\documentclass[12pt,journal,draftclsnofoot,onecolumn,twoside,letterpaper]{IEEEtran}

\usepackage{amsmath, amsthm}
\usepackage{eepic, epic,color}
\usepackage{times}
\usepackage[latin1]{inputenc}
\usepackage{amsfonts}
\usepackage{psfrag}
\usepackage{amsbsy}
\usepackage{amssymb}
\usepackage[mathscr]{eucal}
\usepackage{latexsym}
\usepackage[T1]{fontenc}
\usepackage{amsmath}
\usepackage{comment}
\usepackage[dvips]{graphicx}
\usepackage{bm}
\usepackage{array}
\usepackage{verbatim,mathrsfs, cite}
\usepackage{geometry}
\usepackage{multirow}

\newcounter{lecnum}
\newtheorem{lemma}{Lemma}
\newtheorem{proposition}{Proposition}

\newtheorem{corollary}{Corollary}

\newcommand{\figw}{0.95\columnwidth}
\newcommand{\argmax}{\operatornamewithlimits{argmax}}

\def\B(#1){\hbox{\boldmath$#1$}}
\def\C(#1){{\cal #1}}

\graphicspath{{./Figure/}}

\begin{document}

\title{\LARGE Pricing and Intervention in Slotted-Aloha: \\ Technical Report}

\author{\authorblockN{{\bf Luca Canzian}$^\diamond$, {\bf Yuanzhang Xiao}$^\S$, {\bf Michele Zorzi}$^\diamond$, \\{\bf Mihaela van der Schaar}$^\S$}\\
{$^\diamond$DEI, University of Padova, via Gradenigo 6/B, 35131 Padova, Italy
\\
$^\S$Department of Electrical Engineering, UCLA, Los Angeles CA 90095, USA
}
}

\maketitle

\begin{abstract}
In many wireless communication networks a common channel is shared by multiple users who must compete to gain access to it.
The operation of the network by self-interested and strategic users usually leads to the overuse of the channel resources and to substantial inefficiencies.
Hence, incentive schemes are needed to overcome the inefficiencies of non-cooperative equilibrium.
In this work we consider a slotted-Aloha like random access protocol and two incentive schemes: \emph{pricing} and \emph{intervention}.
We provide some criteria for the designer of the protocol to choose one scheme between them and to design the best policy for the selected scheme, depending on the system parameters.
Our results show that intervention can achieve the maximum efficiency in the \emph{perfect monitoring} scenario. 
In the \emph{imperfect monitoring} scenario, instead, the performance of the system depends on the information held by the different entities and, in some cases, there exists a threshold for the number of users such that, for a number of users lower than the threshold, intervention outperforms pricing, whereas, for a number of users higher than the threshold pricing outperforms intervention.
\end{abstract}


\section{Introduction}\label{sec:intro}

In wireless communication networks, multiple users often share a common channel and contend for access. 
Many distributed Medium Access Control (MAC) protocols, some of them being used in current international standards (e.g., IEEE 802.11 a/b/g/n), have been designed assuming that users are compliant with the protocol rules. 
Unfortunately, a self-interested and strategic user might manipulate the protocol in order to obtain a larger share of the channel resource at the expense of that of the other users.\footnote{In \cite{Bianchi} the 802.11 MAC protocol of a commercial Broadcom chipset is replaced with a state machine execution engine which allows to program and use the desired MAC protocol. Such a capability of modifying MAC protocols results in our concerns for self-interested users in future wireless networks.} 
In the literature, adopting game theoretic approaches \cite{GTOwen}, it is shown that the presence of self-interested users usually leads to the overuse of the channel resource and to substantial inefficiencies \cite{802, Hubaux, slotAloha}.

We consider a slotted-Aloha like random access protocol, where each user transmits within a slot according to some user-chosen probability. 
Without any further mechanism, self-interested users would implement the \emph{always transmit} strategy, resulting in the network collapse.
To make the network robust to selfish users, it is fundamental to design a scheme that provides to the users the incentives to adopt a better (from the network designer point of view) strategy.

In the past decade a lot of research was dedicated to the development 
of such incentive schemes for slotted-Aloha like random access protocols. 
Some of this research, such as \cite{JinKesidis, WangComaniciu, WangComaniciu_WPC, PavanJenTara,YangKimZhangChiangTan_INFOCOM11},
adopts pricing schemes that charge the users for their resource usage.\footnote{Notice that in the literature pricing schemes may refer also to distributed schemes in which the users are cooperative and \emph{fictitious} prices are used to obtain an efficient distributed algorithm. In our case, we consider strategic and selfish users, thus, to be effective, the pricing scheme requires the users to pay real money.} 
In this way, it is in the self-interest of each user to limit its access probability.
Such pricing schemes may achieve the goal of efficient use of network resources.
However, they suffer from the following drawbacks: (1) the designer has to know how the prices affect the users' utilities to design an efficient scheme; (2) it is not clear what do to with the collected money, unless the network is managed by a profit-making enterprise; (3) a secure infrastructure to collect the money is needed.

Recently, a new incentive scheme, called \emph{intervention}, has been proposed in \cite{ParkMihaela_JSAC} and has been applied to MAC problems \cite{ParkMihaela_EURASIP, ParkMihaela_Gamenets}.
In this scheme, an \emph{intervention device} is placed in the network.
Such a device can monitor the users' behavior and intervene affecting the users' resource usage.
The action of the intervention device depends on the actions of the users.
The intervention device provides the incentives for the users to obey a given access probability rule by threatening \emph{punishments} if users disobey.
Intervention is more robust than pricing because users cannot avoid intervention as long as they use network resources, but they might be able to avoid monetary charges.
The implementation of an intervention scheme requires to place an additional device, i.e., the intervention device, in the network.

Repeated games can also encourage cooperative behaviors \cite{PhanParkMihaela}.
In this case users are forced to take into account how their current actions can influence the future actions of the other users. 
A cooperative behavior is induced by punishing deviating users in the future.
Differently from the previously considered methods, this scheme does not require the presence of a central entity.
However, it requires a repeated interaction among users and the users must keep track of their past observations and be able to detect deviations and to coordinate their actions in order to punish deviating users.
We exclude incentive schemes based on repeated games because of these difficulties. 

In this paper we provide the tools to design pricing and intervention schemes to make a random access protocol robust against strategic users.
As in most of the previous works in pricing and intervention, we consider only \emph{linear intervention} and \emph{linear pricing} schemes, because they are simple to implement and yet efficient enough to achieve high performance (or even optimality in some cases).
Simple rules are important in particular for pricing schemes, because the users might not accept to pay for their resource usage following complex rules.
It is difficult to argue between different incentive schemes in general: depending on the particular deployment scenario, the performance criterion, and the implementation issues, each one of the incentive schemes can be better than the others.
The problem of the network designer is to identify the scheme that best fits its requirements and to design the best policy for the selected scheme.

The complexity of the design process and the performance achievable depend on various features of the system, such as the number of users, the users' heterogeneity, the capability of monitoring the users' actions and the information held by the designer and the users.
To the best of our knowledge, this is the first work that compares intervention and pricing in terms of the network environment, the knowledge of the designer and the knowledge of the users. 
We focus on a simple MAC protocol, slotted-Aloha, because it makes it possible to formulate a simple game in which the outcomes can be computed analytically, to highlight the consequence of not taking into account the strategic nature of some users when designing a MAC protocol, and to obtain important insights about possible solutions to such a problem.
For these features slotted-Aloha is widely used in game theoretic studies \cite{JinKesidis, WangComaniciu, WangComaniciu_WPC, PavanJenTara,YangKimZhangChiangTan_INFOCOM11, ParkMihaela_EURASIP, ParkMihaela_Gamenets}.
The extension of this paper to more realistic MAC protocols will be considered in future works.

This paper is divided into two main parts.
In the first part, we consider the \emph{perfect monitoring} scenario, i.e., we assume that the users' actions are estimated without errors.
We show that intervention can achieve the maximum efficiency, i.e., the maximum social welfare, while pricing is able to reach an efficient use of the network resources but the positive payments subtracted from the users' utilities prevent it to achieve the maximum social welfare

In the second part, we consider an \emph{imperfect monitoring} scenario, assuming that a uniformly distributed noise term is added to the estimated actions.
We derive the optimal pricing and intervention schemes and quantify the performance achievable in this scenario, assuming that (1) neither the designer nor the users are aware of the estimation errors (i.e., they believe that the designer is able to observe the users' actions perfectly), (2) only the designer is aware of the estimation errors, and (3) both the designer and the users are aware of the estimation errors.
In the imperfect monitoring scenario, the performance of the intervention scheme degrades considerably as the number of users increases and the information held by the designer and the users plays an important role.
In particular, for case (3) there exists a threshold for the number of users such that, for a number of users lower than the threshold, the intervention scheme outperforms the pricing scheme, while for a number of users higher than the threshold the pricing scheme outperforms the intervention scheme.
In the other cases intervention allows to obtain higher performance than pricing.

Table \ref{tab:1} summarizes some results obtained and the implementation requirements for the considered schemes.
The analysis in this paper can serve as a guideline for a designer to select between pricing and intervention and to design the best policy for the selected scheme, depending on some system parameters such as the number of users, the statistics of the monitoring noise and the information held by the designer and the users.

Despite its practical importance, very few works address the impact of the monitoring errors and the information heterogeneity on the design and performance of an incentive scheme.
To the best of our knowledge, no prior work on pricing considers the issue of imperfect monitoring on users' actions.
As to the intervention scheme, both \cite{ParkMihaela_EURASIP} and \cite{ParkMihaela_Gamenets} consider the imperfect monitoring scenario. 
\cite{ParkMihaela_EURASIP} adopts the same noise model we use, but it simplifies the analysis limiting the users' action space, whereas \cite{ParkMihaela_Gamenets} considers a different type of imperfect monitoring, whose distribution depends on the length of the time the intervention device takes to estimate users' actions.
However, in both works it is assumed that the designer and the users are aware of the imperfect monitoring model.
In our work we analyze the effect of the information heterogeneity, considering also the cases in which nobody is aware of the estimation errors and in which only the designer is aware of the estimation errors.
This provides understanding on how robust the considered incentive schemes are with respect to the heterogeneity of information.

The remainder of this paper is organized as follows.
In Section \ref{sec:sys} we describe the considered MAC protocol.
We introduce the games that model the interaction between strategic users and we formulate the problem of designing efficient incentive schemes in Section \ref{sec:game}.
In Section \ref{sec:per_mon} we derive the optimal pricing and intervention schemes to adopt in the perfect monitoring scenario and we quantify the performance achievable. 
We consider the imperfect monitoring scenario in Section \ref{sec:imp_mon}, for three different cases, depending on who is aware of the imperfect monitoring model.
Section \ref{sec:con} concludes with some remarks.

\begin{table}
\begin{center}
\begin{tabular}{|l|m{2.6cm}|m{3cm}|m{3.5cm}|m{3.1cm}|}
\hline
\multirow{2}{*}{\textbf{Incentive scheme}} & \multicolumn{2}{|c|}{\textbf{Performance}} & \multicolumn{2}{|c|}{\textbf{Implementation challenges}} \\ \cline{2-5}
& \textbf{Perfect monitoring} & \textbf{Imperfect monitoring} & \textbf{Infrastructure \newline requirements} & \textbf{Information \newline requirements} \\
\hline
None & Network collapse & Network collapse & None & None \\
\hline
Pricing & Suboptimal & Suboptimal \newline Better than intervention if the number of users is large & Device that monitor users' resource usage; secure and reliable method to collect the payments & Designer needs users' utilities \newline Users need unit prices \\
\hline
Intervention & Optimal & Suboptimal \newline Better than pricing if the number of users is small & Device that monitors users' actions and intervenes & Designer needs users' utilities \newline Users need intervention rule\\ 
\hline
\end{tabular}
\end{center}
\caption{Pricing and intervention for the considered random access protocol: performance and requirements.}
\label{tab:1}
\end{table}

\section{System model}\label{sec:sys}

We consider a wireless network of $n$ users that share a common channel and we make the following assumptions for the contention model:
\begin{itemize}
\item Time is slotted and slots are synchronized;
\item Users always have packets to transmit in every slot;
\item If a packet is received, the receiver immediately sends an acknowledgment (ACK) packet;
\item The transmission of a packet and the (possible) corresponding ACK is completed within a slot;
\item A packet is received successfully if and only if it does not collide with other transmissions;
\item Each user $i$ selects a transmission probability $p_i \in \left[ 0 \, 1 \right] $ at the beginning of the communication and will transmit with the same probability $p_i$ in every time slot, i.e., there are no adjustments in the transmission probabilities. This excludes coordination among users, for example, using time division multiplexing.
\end{itemize}
Notice that ACK packets are always successfully received because they are transmitted over idle channels. 

Denoting with $p = \left( p_1, \ldots, p_n \right) $ the transmission probability vector, the average throughput (in packets per slot) of user $i$ is given by
\begin{equation}
T_i(p) = p_i \prod_{j=1 , j \neq i}^n (1-p_j)
\end{equation}
The resource usage of user $i$ is therefore proportional to $i$'s transmission probability.

We assume that the utility of user $i$ is given by 
\begin{equation}
U_i(p) = \theta_i \ln T_i(p) 
\label{eq:Ui}
\end{equation}
where the parameter $\theta_i > 0$ allows to differentiate between different classes of users. 
The higher $\theta_i$, the higher user $i$'s valuation for the throughput.
The logarithm makes the utility a concave function, 
which models the fact that the users usually have more desire to increase their own throughput when it is low than when it is high.

We define the social welfare of the network as the sum of all users' utilities:
\begin{equation}
U(p) = \sum_{i=1}^n U_i(p) 
\label{eq:U}
\end{equation}

Finally, the network is said to operate optimally if the users choose the transmission probabilities that maximize Eq. (\ref{eq:U}).
It is straightforward to check that the Hessian of $U(p)$ is a diagonal matrix with strictly negative diagonal entries, therefore it is negative definite. 
Imposing the partial derivatives equal to $0$, the unique transmission probability vector $p^* = \left( p_1^*, \ldots, p_n^* \right) $ that maximizes Eq. (\ref{eq:U}) is given by 
\begin{equation}
p_k^* = \dfrac{\theta_k}{\sum_{i=1}^n \theta_i} \;\;\; , \;\;\; k = 1, \dots, n
\label{eq:p_opt}
\end{equation}
We say that $U(p^*)$ is the maximum efficiency utility.

In order to adopt the optimal transmission probability, the users need to know the sum of the valuations $\theta_i$ of the other users.
This information must be spread in the network at the beginning of the communication.
This can be done either in a distributed way or in a centralized way. 
In particular, in the last case an entity (e.g., a predetermined user or the access point) might collect the users' valuations and broadcast to all users the value $\sum_{i=1}^n \theta_i$.
Once the users have this information, they can locally compute their optimal transmission probabilities according to Eq. (\ref{eq:p_opt}) and adopt them.

\section{Game model and design problem formulation}\label{sec:game}

While the network optimal transmission policy $p^*$ is easy to compute, the actual transmission probability selected by each user depends on the objective of that user.
If the users are compliant with the optimal policy, then they compute and adopt $p^*$ and the network operates optimally.
However, if the users are self-interested and strategic, instead of complying with the optimal policy they will adopt the transmission probabilities that optimize their own utility. 
Since the interests of individual users are different from the interests of the group of users as a whole, the network might (and usually will) operate inefficiently.

To analyze the interaction between strategic decision-makers, we exploit the models offered by game theory.
We define the contention game
\begin{equation}
\Gamma = \left( \mathcal{N}, A, \left\lbrace U_i(\cdot) \right\rbrace_{i=1}^n  \right)
\label{eq:gamma}
\end{equation}
where $\mathcal{N} = \left\lbrace 1, 2, \cdots, n \right\rbrace $ denotes the set of users, $A = \times_{i=1}^n \left[ 0 , \, 1 \right]^n $ denotes the action space and $U_i: A \rightarrow \Re$ is the utility of a generic user $i$, defined by Eq. (\ref{eq:Ui}).
The action for user $i$ represents the transmission probability $p_i$ chosen by user $i$. 
Throughout the paper, we will use the terms action and transmission probability interchangeably, and similarly for action profile (a collection of the users' actions) and transmission probability vector.

A widely accepted solution concept for non-cooperative games is the Nash Equilibrium ($NE$), defined as the action profile $p^{NE}$ so that each user obtains its maximum utility given the actions of the other users, i.e., 
\begin{equation}
U_i \left( p^{NE} \right) \geq U_i \left( p_i, p_{-i}^{NE} \right) \;\; , \;\; \forall \, i \in \mathcal{N} \; , \; \forall \, p_i \in \left[ 0 , \, 1\right] 
\end{equation} 
where $p_{-i}$ denotes the transmission probabilities of all the users except for user $i$.

The $NEs$ of the contention game $\Gamma$ can be easily characterized considering the following cases.
\begin{itemize}
\item[1)] Assume that all user, except for user $i$, adopt a transmission probability strictly smaller than $1$.
Then the utility of user $i$ is increasing in $p_i$: the higher the transmission probability chosen by $i$ the higher $i$'s throughput. Thus, $i$ chooses $p_i=1$.
\item[2)] Assume that there is at least a user $j \neq i$ that adopts a transmission probability equal to $1$. Then the channel is always busy and user $i$ obtains a throughput equal to $0$, regardless of its transmission probability. 
\end{itemize}
In case 1) $p_i=1$, in case 2) $p_j=1$
Thus, $p$ is a Nash Equilibrium of the contention game $\Gamma$ if and only if at least one user adopts a transmission probability equal to $1$.
Notice that $p_i = 1$ is a \emph{weakly dominant strategy} for every user $i$, i.e., $u_i(1, p_{-i}) \geq u_i(p)$, for every action profile $p$.
In our contention game each user has an incentive to adopt the \emph{always transmit} strategy, resulting in network collapse.

Here we ask if it is possible to design the network to make it robust against strategic users. 
We want to introduce some mechanism to deter the users from adopting high transmission probabilities.
The incentive schemes we consider belong to two classes:
\begin{itemize}
\item Pricing: users are charged depending on their transmission probabilities
\item Intervention: the users' resource usage is affected by the \emph{intervention device}, in a way that depends on the users' transmission probabilities
\end{itemize}

The interaction between the designer, the users and the system can be roughly summarized into three stages, (1) the design stage, (2) the information exchange stage, and (3) the transmission stage.

In the \emph{design stage} the designer designs the pricing or intervention scheme.
Specifically, the designer predicts strategic users' actions given any pricing or intervention scheme, and chooses the pricing or intervention scheme that results in the most desired outcome.
This is done once, then the designer leaves the system forever.
Notice that, to efficiently design these schemes, the designer has to know how pricing or intervention affect the users' utilities. 
This might be easier for the intervention scheme, in which the users' throughput is altered.
In this case the designer has to know only the relation between the throughput and the utility of each user.
Differently, in the pricing schemes users are charged for their resource usage. 
Hence, the designer has to know how throughput and payments are connected to the utility of each user.
In this work we implicitly assume that the designer knows these dependencies, because we focus on a particular relation between the utilities, the throughput, and the payments.

In the \emph{information exchange stage} some useful information is collect and, possibly, distributed.
The intervention device (or the device that manages the payments in the pricing scheme) has to identify the users that are connected to the network, has to inform them about the adopted intervention or pricing scheme, and has to learn the action they select.
For the latter point, 
it can estimate them through its \emph{monitoring technology}, e.g., by counting the number of correct transmissions of each user in a certain time interval.
Since this time interval must be finite, the estimation might be affected by errors. 
To consider the impact of this imperfect estimation we will denote by $\hat{p}_i$ the estimated action of user $i$, by $\hat{p}$ the estimated action profile and by $\pi_i(\hat{p}_i \mid p_i)$ the probability density function of $i$'s estimated action, given that $i$'s action is $p_i$.
We say that the monitoring is \emph{perfect} if the users' actions are estimated without errors, i.e., $\hat{p}_i$ coincides with $p_i$.\footnote{In this case $\pi_i(\hat{p}_i \mid p_i)$ might be thought as a Dirac delta function centered in $p_i$.}
We say that the monitoring is \emph{imperfect} if the estimates are affected by errors, i.e., there is a positive probability that $\hat{p}_i$ is different from $p_i$.

In the \emph{transmission stage} the users transmit the packets adopting the same transmission probability and, in the meantime, they have to pay for their resource usage based on the pricing scheme, or their resource usage is affected based on the intervention scheme.

In this paper we play the role of a benevolent designer that seeks to design the pricing and intervention rules to maximize the social welfare of the system in the transmission stage.
We neglect the social welfare obtained in the information exchange stage because we assume that the transmission stage length is much longer than that of the information exchange stage.

\subsection{Pricing}\label{sec:game1}

Pricing schemes use monetary charges to deter users' greediness.
If $i$'s payment is increasing in $i$'s resource usage, user $i$ might find it convenient to limit its transmission probability.
In general, user $i$ is charged according to the \emph{pricing rule} $f_i^P: \left[ 0 ,\, 1\right] \rightarrow \Re $, which is a function of $i$'s estimated action $\hat{p}_i$.
Assuming that the payments affect additively the users' utilities, $i$'s expected utility is given by
\begin{equation}
U_i^P(p) = \mathbb{E} \left[ \theta_i \ln T_i(p) - f_i^P(\hat{p}_i) \right] = \theta_i \ln T_i(p) - \int_0^1 \pi_i(\hat{p}_i \mid p_i) f_i^P(\hat{p}_i) \partial \hat{p}_i
\label{eq:Ui_P}
\end{equation}
where $\mathbb{E} \left[ \cdot \right]$ is the expectation operator.

Once a pricing scheme is selected and communicated to the users, the interaction among users can be modeled through the game 
\begin{equation}
\Gamma^P = \left( \mathcal{N}, A, \left\lbrace U_i^P(\cdot) \right\rbrace_{i=1}^n  \right)
\label{eq:gamma_P}
\end{equation}

Among all the possible pricing rules, there is one class of rules that is particularly interesting, namely, the class of \emph{linear pricing rules}, in which users are charged linearly with respect to their transmission probabilities, i.e.,
\begin{equation}
f^P_i(\hat{p}_i) = c_i \hat{p}_i
\label{eq:f_P}
\end{equation}
where $c_i \geq 0$ is the unit price.
We restrict our attention to the linear pricing rules, as done in most of the pricing literature, because they are computationally simple to implement and we do not lose much, in term of performance, in doing so.

Once the prices $c = \left( c_1, \ldots, c_n \right)$ are fixed, since we will prove the existence and uniqueness of the $NE$ of the game $\Gamma^P$, the social welfare can be uniquely determined.
The goal of the designer is to choose the unit prices $c = \left( c_1, \ldots, c_n \right)$ to maximize the social welfare, i.e., it has to solve the following Pricing Design (\textbf{PD}) problem:
\begin{align}
\textbf{PD} \;\;\;\;\; & \argmax_{c} \sum_{i \in \mathcal{N}} U_i^P(p^{NE}) \nonumber\\
& \mbox{subject to:} \nonumber\\
& c_i \geq 0 \;\; , \;\; \forall \, i \in \mathcal{N} \nonumber\\
& U_i^P(p^{NE}) \geq U_i^P(p_i, p_{-i}^{NE}) \;\; , \;\;  \forall \, p_i \in \left[ 0 ,\, 1\right] \;\; , \;\; \forall \, i \in \mathcal{N} \nonumber
\end{align}

\subsection{Intervention}\label{sec:game2}

In the intervention framework the designer deploys in the network an intervention device that monitors the users' actions and can intervene adopting itself an action that affects the users' resource usage.
In our case, we assume that the intervention device is able to correctly recognize the packets transmitted by different users and to estimate the users' actions.
If the packet of a generic user $i$ is correctly received, the intervention device may choose to jam its ACK\footnote{Many works on security, such as \cite{Milcom06, PerJam, ProLa}, take into consideration the possibility of performing intelligent jamming in which the jamming signal is concentrated on control packets.} depending on the estimate of its action. Specifically, the intervention device jams the ACK sent to user $i$ with a probability that is given by the \emph{intervention rule} $f_i^I: \left[ 0 ,\, 1\right] \rightarrow \left[ 0 ,\, 1\right] $, which is a function of the estimated action $\hat{p}_i$.

The intervention level $f_i^I (\hat{p}_i)$ must be interpreted as a \emph{punishment} to user $i$ after having deviated from a recommended (socially-beneficial) action.
Such punishments are a threat to users, and must be designed such that the users find in their self-interest to adopt the recommended actions. 
At the same time, when users adopt the recommended actions, the intervention level must be minimized (possibly, nullified), to avoid to decrease the users' utilities.

Different from pricing, intervention changes the structure of the utility of each user affecting directly their resource usage. 
In fact, the average throughput of user $i$ is now given by
\begin{align}
T_i^I(p) &= \mathbb{E} \left[ p_i \left( 1 - f_i^I(\hat{p}_i) \right) \prod_{j=1 , j \neq i}^n (1-p_j) \right] = p_i \left( 1 - \int_0^1 \pi_i(\hat{p}_i \mid p_i) f_i^I(\hat{p}_i) \partial \hat{p}_i \right) \prod_{j=1 , j \neq i}^n (1-p_j) 
\label{eq:Ti_I}
\end{align}
where $\int_0^1 \pi_i(\hat{p}_i \mid p_i) f_i^I(\hat{p}_i) \partial \hat{p}_i$ represents the average intervention level.

The utility of user $i$ is modified accordingly
\begin{equation}
U_i^I(p) = \theta_i \ln T_i^I(p)
\label{eq:Ui_I}
\end{equation} 

Once the intervention rules are selected and communicated to the users, the interaction between the users can be modeled through the game 
\begin{equation}
\Gamma^I = \left( \mathcal{N}, A, \left\lbrace U_i^I(\cdot) \right\rbrace_{i=1}^n  \right)
\label{eq:gamma_I}
\end{equation}
We say that the intervention rules $f^I = \left( f_1^I, \ldots, f_n^I \right) $ \emph{sustain} an action profile $p$, if $p$ is a $NE$ of $\Gamma^I$.

Among all the possible intervention rules, there is one class of rules that is particularly interesting, namely, the class of \emph{affine intervention rules}.
$f_i^I: \left[ 0 ,\, 1\right] \rightarrow \left[ 0 ,\, 1\right] $ is an affine intervention rule if 
\begin{align}
f_i^I(\hat{p}_i) = \left[ r_i (\hat{p}_i - \tilde{p}_i) \right]_0^1
\label{eq:int_rule}
\end{align}
for certain parameters $\tilde{p}_i \in \left[ 0 ,\, 1\right]$ and $r_i \geq 0$, where $\left[ \cdot \right]_a^b = \min \left\lbrace \max \left\lbrace a, \cdot \right\rbrace, b \right\rbrace $.

In an affine intervention rule, $\tilde{p}_i$ represents a target action for user $i$ while $r_i$ represents the rate of increase of the intervention level due to an increase in $i$'s action.
If the estimated action $\hat{p}_i$ is lower than or equal to the target action $\tilde{p}_i$, then the intervention level is equal to $0$.
If the estimated action $\hat{p}_i$ is higher than the target action $\tilde{p}_i$, then the intervention level is proportional to $\hat{p}_i - \tilde{p}_i$, until it saturates to $1$.

For $r_i \rightarrow + \infty$, the intervention device jams the ACKs sent to user $i$ whenever it detects that $i$ is adopting an action higher than the target one.
Such a rule, which we refer to as an \emph{extreme rule}, represents the strongest punishment that the intervention device can adopt.

We restrict our attention to the affine intervention rules because they are computationally simple to implement and we do not lose much, in term of performance, in doing so (as we will see, in some cases such rules are even able to achieve the benchmark optimum).

Once the parameters $\tilde{p} = \left( \tilde{p}_1, \ldots, \tilde{p}_n \right)$ and $r = \left( r_1, \ldots, r_n \right)$ are fixed, and assuming that the users coordinate to the best (from the social welfare point of view) $NE$ of the game $\Gamma^I$ \footnote{The existence of $NEs$ will be proved for the considered scenarios and it is easy to coordinate the users to the best $NE$. In fact, we will prove that the best $NE$ is uniquely determined by $\tilde{p}$.}, the social welfare can be determined.
The goal of the designer is to choose the parameters $\tilde{p}$ and $r$ to maximize the social welfare, i.e., it has to solve the following Intervention Design (\textbf{ID}) problem:
\begin{align}
\textbf{ID} \;\;\;\;\; & \argmax_{\tilde{p}, r} \left[ \max_{p^{NE}} \sum_{i \in \mathcal{N}} U_i^I(p^{NE}) \right] \nonumber\\
& \mbox{subject to:} \nonumber\\
& \tilde{p}_i \in \left[ 0 ,\, 1\right] \;\; , \;\; r_i \geq 0 \;\; , \;\; \forall \, i \in \mathcal{N} \nonumber\\
& U_i^I(p^{NE}) \geq U_i^I(p_i, p_{-i}^{NE}) \;\; , \;\;  \forall \, p_i \in \left[ 0 ,\, 1\right] \;\; , \;\; \forall \, i \in \mathcal{N} \nonumber
\end{align}
Differently from the \textbf{PD} problem, the \textbf{ID} problem requires a maximization with respect to the $NEs$ because of the non uniqueness of the $NE$.

\section{Perfect monitoring}\label{sec:per_mon}

In this section we assume that the estimated actions are equal to the real actions, i.e., $\hat{p}_i = p_i$, for every user $i \in \mathcal{N}$. 
Hence, in Eq. (\ref{eq:Ui_P}) and (\ref{eq:Ti_I}) the integrals must be substituted, respectively, with $f_i^P(p_i)$ and $f_i^I(p_i)$.
In the following we compute the optimal linear pricing scheme and affine intervention rule that a designer should adopt to maximize
the social welfare if the monitoring is perfect.

\subsection{Pricing design}\label{sec:per_pri}

Given a linear pricing scheme $c_i$, $i \in \mathcal{N}$, the interaction between users in the perfect monitoring scenario adopting pricing is modeled with the game 
\begin{equation}
\Gamma^P = \left( \mathcal{N}, A, \left\lbrace U_i^P(\cdot) \right\rbrace_{i=1}^n  \right)
\label{eq:gamma_P2}
\end{equation} 
where 
\begin{equation}
U_i^P(p) = \theta_i \ln \left[ p_i \prod_{j=1 , j \neq i}^n (1-p_j) \right] - c_i p_i
\label{eq:Ui_P2}
\end{equation}
The goal of the designer is to design the unit prices $c$ to maximize the social welfare in the presence of strategic users, solving the \textbf{PD} problem with the utilities given by Eq. (\ref{eq:Ui_P2}).

\begin{lemma}
The unique $NE$ of the game $\Gamma^P$ is $p_k^{NE} = \dfrac{\theta_k}{c_k}$, $k \in \mathcal{N}$.
\label{lem:1}
\end{lemma}

\begin{IEEEproof}

To compute the best response function of users $k$, we use the first order condition.
First, we check that $U_k^P(p)$ is concave in $p_k$ (i.e., the second derivative with respect to $p_k$ is negative). 
Then, we set to $0$ the first derivative of $U_k^P(p)$, with respect to $p_k$.
\begin{align}
\frac{\partial U_k^P(p)}{\partial p_k} &= \dfrac{\theta_k}{p_k} - c_k \nonumber\\
\frac{\partial^2 U_k^P(p)}{\partial p_k^2} &= - \dfrac{\theta_k}{p_k^2} < 0 \nonumber\\
\frac{\partial U_k^P(p)}{\partial p_k} &= 0 \; \longrightarrow \; p_k =  \dfrac{\theta_k}{c_k} 
\end{align}
\end{IEEEproof}

\begin{proposition}
The optimal pricing scheme to adopt is $c_k^* = \sum_i \theta_i $.
\end{proposition}

\begin{IEEEproof}

We want to find the unit prices $c_k$, $k \in \mathcal{N}$, so that the social welfare $U(p) = \sum_{i=1}^n U_i^P(p) $ is maximized, assuming that the users adopt the $NE$ action profile (i.e., we have to substitute $c_k$ with $\dfrac{\theta_k}{p_k}$ into the expression of $U(p)$).
We first prove that $U(p)$ is a (multivariable) concave function, by checking its Hessian.
\begin{align}
\frac{\partial U(p)}{\partial p_k} &= \dfrac{\theta_k}{p_k} - \dfrac{\sum_{i\neq k} \theta_i}{1-p_k} \nonumber\\
\frac{\partial ^2 U(p)}{\partial p_k^2} &= - \dfrac{\theta_k}{p_k^2} - \dfrac{\sum_{i\neq k} \theta_i}{(1-p_k)^2} < 0\nonumber\\
\frac{\partial ^2 U(p)}{\partial p_k dp_i} &= 0 \; , \; i \neq k 
\end{align}
The Hessian of $U(p)$ is negative definite (it is a diagonal matrix with strictly negative diagonal entries), so $U(p)$ is concave.
Thus, the global maximizer of $U(p)$ can be obtained with the first order condition
\begin{align}
\frac{\partial U(p)}{\partial p_k} &= 0 \; \longrightarrow \; p_k =  \dfrac{\theta_k}{\sum_i \theta_i} \; \longrightarrow \; c_k = \sum_i \theta_i \;\; , \;\; k\in \mathcal{N}
\end{align}

\end{IEEEproof}

Notice that the transmission probabilities adopted by the users in the optimal pricing policy are equal to the transmission probabilities adopted by compliant users to maximize the social welfare, i.e., $p_k^{NE} = \dfrac{\theta_k}{c_k^*} = p^*$, where $p^*$ is defined in Eq. (\ref{eq:p_opt}).

\subsection{Intervention design}\label{sec:per_int}

Given an affine intervention rule  $r_i$ and $\tilde{p}_i$, $i \in \mathcal{N}$, the interaction between users in the perfect monitoring scenario adopting intervention is modeled with the game 
\begin{equation}
\Gamma^I = \left( \mathcal{N}, A, \left\lbrace U_i^I(\cdot) \right\rbrace_{i=1}^n  \right)
\label{eq:gamma_I2}
\end{equation} 
where 
\begin{equation}
U_i^I(p) = \theta_i \ln \left[ p_i \left( 1 -  \left[ r_i (p_i - \tilde{p}_i) \right]_0^1 \right) \prod_{j=1 , j \neq i}^n (1-p_j) \right] 
\label{eq:Ui_I2}
\end{equation}
The goal of the designer is to design the intervention rule to maximize the social welfare in the presence of strategic users, solving the \textbf{ID} problem with the utilities given by Eq. (\ref{eq:Ui_I2}).

Notice the $\tilde{p}_k = 1$ and $\tilde{p}_k = 0$ represent trivial cases.
If $\tilde{p}_k = 1$ the intervention device never jams the ACK sent to user $k$, $\forall \, p_k$, and in this case $p_k = 1$ represents a weakly dominant strategy, as discussed in Section \ref{sec:game}.
If $\tilde{p}_k = 0$ user $k$ is punished whenever it transmits with positive probability. 
However, the aim of the designer is to maximize the social welfare, hence, it must first guarantee a positive throughput to every user. 
Thus, it is always more beneficial to consider a $\tilde{p}_k$ slightly higher than $0$ instead of $0$. 
For this reason, in the following we focus on intervention rules in which $\tilde{p}_k \in \left( 0, 1\right) $, $\forall \, k$.

\begin{lemma}
$\tilde{p} = \left( \tilde{p}_1 , \ldots, \tilde{p}_n \right) $ is a $NE$ of the game $\Gamma^I$ if and only if $r_k \geq \dfrac{1}{\tilde{p}_k}$, for every user $k \in \mathcal{N}$.
Moreover, once $\tilde{p}$ and $r_k \geq \dfrac{1}{\tilde{p}_k}$ are fixed, among all the $NE$s of $\Gamma^I$,  $\tilde{p}$ is (individually and socially) the best. 
\label{lem:2}
\end{lemma}

\begin{IEEEproof}
We can write $r_k = \dfrac{1}{\tilde{p}_k + \delta}$, for some constant $\delta > -\tilde{p}_k$. 
Then
\begin{equation}
U_k^I(p_k, \tilde{p}_{-k}) =
\left\lbrace
\begin{array}{ll}
\theta_k \ln \left[ p_k \prod_{j \neq k} (1-\tilde{p}_j) \right] & \mbox{if} \;\; p_k < \tilde{p}_k \\
\theta_k \ln \left[ \dfrac{ -p_k^2 + 2 \tilde{p}_k p_k + \delta p_k}{\tilde{p}_k + \delta} \prod_{j \neq k} (1-\tilde{p}_j) \right] & \mbox{if} \;\; \tilde{p}_k \leq p_k \leq 2 \tilde{p}_k + \delta \\
-\infty & \mbox{if} \;\; p_k > 2 \tilde{p}_k + \delta
\end{array}
\right. 
\end{equation}

We study the sign of $\dfrac{\partial U_k^I(p_k, \tilde{p}_{-k})}{\partial p_k}$ in the interval $\left[ 0,  2 \tilde{p}_k + \delta \right]$ to obtain the best action for user $k$.
\begin{equation}
\dfrac{\partial U_k^I(p_k, \tilde{p}_{-k})}{\partial p_k} =
\left\lbrace
\begin{array}{ll}
\dfrac{\theta_k}{p_k} & \mbox{if} \;\; p_k < \tilde{p}_k \\
\theta_k \dfrac{2 \left( \tilde{p}_k - p_k \right) + \delta}{p_k \left( 2 \tilde{p}_k - p_k + \delta \right) } & \mbox{if} \;\; \tilde{p}_k \leq p_k \leq 2 \tilde{p}_k + \delta \\
\end{array}
\right. 
\end{equation}

If $\delta \leq 0$ (i.e, $r_k \geq \dfrac{1}{\tilde{p}_k}$), $U_k^I(p_k, \tilde{p}_{-k})$ is continuous, increasing in $p_k$ for $p_k < \tilde{p}_k$ and decreasing otherwise.
Thus, $\tilde{p}_k$ is the best action for user $k$.

If $\delta > 0$ (i.e, $r_k < \dfrac{1}{\tilde{p}_k}$), $U_k^I(p_k, \tilde{p}_{-k})$ is continuous, increasing in $p_k$ for $p_k < \tilde{p}_k + \dfrac{\delta}{2} $ and decreasing otherwise.
Thus, $\tilde{p}_k + \dfrac{\delta}{2}$ ($> \tilde{p}_k$) is the best action for user $k$.

Hence, $\tilde{p}$ is a $NE$ if and only if $r_k \geq \dfrac{1}{\tilde{p}_k}$, $ \forall \, k$.
Notice also that, in this situation, $\tilde{p}$ is a weakly dominant strategy: it is in the self interest of each user $k$ to adopt $\tilde{p}_k$, independently of the strategies of the other users.
Thus, the users will coordinate to such $NE$.

Finally, notice that other $NE$s of $\Gamma^I$ can only be obtained when at least two users transmit with probability $1$.
In fact, in this situation no user can increase its utility changing its action.
Actually, the utility can not decrease either: it is constant and it is the worst (individually and socially) possible utility, corresponding to the situation in which the throughput of each user is equal to $0$.
\end{IEEEproof}

\begin{proposition}
The optimal affine intervention rule to adopt is $r_k \geq \dfrac{1}{p_k^*}$ and $\tilde{p}_k = p_k^*$, for every user $k$, where $p_k^*$ is defined in Eq. (\ref{eq:p_opt}).
\label{prop:1}
\end{proposition}

\begin{IEEEproof}
Given the actions of the users, the utility of a user and the social welfare are decreasing as the intervention level for that user increases.
However, using the intervention rule $r_k \geq \dfrac{1}{p_k^*}$ and $\tilde{p}_k = p_k^*$, $\forall \, k$, the users have the incentive to adopt the action profile $p = p^*$ and, at the same time, the intervention level they are subjected to is equal to $0$.
Thus, the outcome of the system is equal to the benchmark optimum.
Finally, this implies that $r_k \geq \dfrac{1}{p_k^*}$ and $\tilde{p}_k = p_k^*$ defines an optimal affine intervention rule, and, more specifically, it defines also an optimal intervention rule within the class of all intervention rules.
\end{IEEEproof}

\begin{corollary}
The optimal affine intervention rule is optimal in the class of all intervention rules.
\label{cor:1}
\end{corollary}

\subsection{Comparison between pricing and intervention and some results}\label{sec:com}

By adopting either pricing or intervention the designer can provide the incentive for strategic users to choose the optimal action profile of Eq. (\ref{eq:p_opt}).
The efficiency of the utilization of the channel resource is optimized with respect to the valuations $\theta_i$, $i \in \mathcal{N}$, of the users.
However, there is a big difference between pricing and intervention.
Intervention schemes reach this objective by threatening the users to intervene if they do not follow the recommendations, although at the equilibrium the intervention is not triggered and therefore the resource usage is not affected.
Conversely, pricing schemes charge each user that transmits with a positive probability, thus affecting its utility and the social welfare.
Hence, only the intervention scheme is able to achieve the optimal social welfare that can be obtained when users behave cooperatively, i.e., when they comply to a prescribed protocol that maximizes the social welfare.

\begin{figure}
     \centering
          \includegraphics[width=\figw]{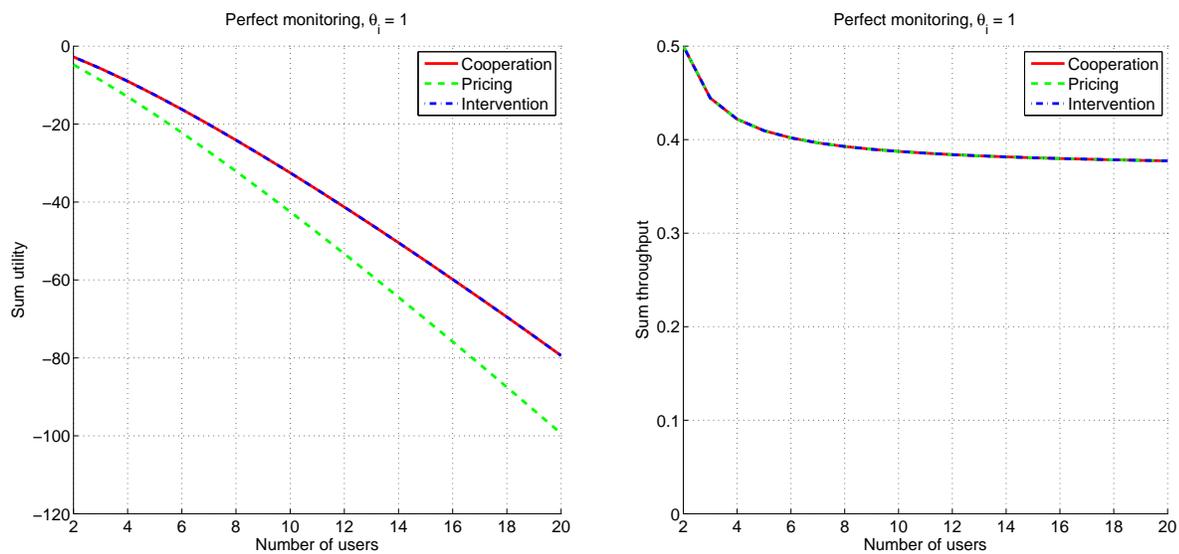}
\caption{Social welfare and total throughput vs. number of users, in the perfect monitoring scenario}
\label{fig:1}
\end{figure}

In Fig. \ref{fig:1} the social welfare and the total throughput in the perfect monitoring scenario are plotted as a function of the number of users in the system, both assuming that the users behave cooperatively, and adopting the pricing and intervention schemes derived in Sections \ref{sec:per_pri} and \ref{sec:per_int} to enforce the users' actions.
A symmetric case is considered, i.e., $\theta_i = 1$, $\forall \, i \in \mathcal{N}$.
Thus, the optimal transmission policy in the cooperative scenario, defined by Eq. (\ref{eq:p_opt}), is $p_k^* = \dfrac{1}{n}$, for every user $k$.

The results confirm the above discussion: both schemes are able to obtain the same total throughput of the cooperative case, but only the intervention scheme is able to maximize the (total) users' satisfaction.
In fact, there is a finite gap, which increases as the number of users increases, between the optimal social welfare and the one achievable with the pricing scheme. 
Finally, notice that the social welfare always decreases as the number of users increases because there are more collisions and the number of unexploited slots increases, resulting in an inefficient utilization of the channel; this is an unavoidable consequence of the lack of coordination.

\section{Imperfect monitoring}\label{sec:imp_mon}
We now study whether the qualitative results obtained for the perfect monitoring scenario still hold for the imperfect monitoring case. 
In this section we will see that there is a substantial difference for the intervention scheme when the monitoring is imperfect.
The intuition behind it is related to the possibility that the estimation errors trigger the intervention even though the users are adopting the recommended actions.
As for the pricing scheme, if the expectations of the estimated actions are equal to the real actions, each user might be overcharged or undercharged. 
On average, it is charged correctly, therefore the performance is not strongly affected.

The imperfect monitoring model we consider for the estimation of user $i$'s action is an additive noise term that is uniformly distributed in $\left[ -\epsilon_i, \, \epsilon_i \right] $, with $0 < \epsilon_i \ll 1$, i.e.,
\begin{align}
\hat{p}_i &= [p_i + \textit{n}_i]_0^1 \nonumber\\
\textit{n}_i &\sim \mathcal{U} \left[ -\epsilon_i, \, \epsilon_i \right] 
\end{align}

In the following we compute the best linear pricing scheme and affine intervention rule that a designer should adopt to maximize
the social welfare for different scenarios, depending on the information that the designer and the users have about the imperfect monitoring.
In particular, we consider the following cases:
\begin{itemize}
\item[1)] Nobody is aware of the estimation errors: neither the designer nor the users know about the existence of the noise, and think that the designer can estimate perfectly the users' actions.
\item[2)] The designer is aware of the estimation errors: the designer knows about the existence and the distribution of the noise, while the users think that the designer can estimate perfectly their actions.
\item[3)] Everybody is aware of the estimation errors: both the designer and the users know about the existence and the distribution of the noise.
\end{itemize}

\subsection{Nobody is aware of the estimation errors}\label{sec:imp_mon1}

In this scenario both the designer and the users believe that the users' estimated actions, $\hat{p}$, are equal to the real ones, $p$.
The additive noise $\textit{n}_i$ might be caused by a physical phenomenon which is not predicted by the designer and the users. 
As an example, one component of the monitoring technology might have, at a certain point, a malfunctioning that is not revealed and introduces noise in the measurements.

Both the designer and the users have a wrong perception of the reality: they both believe that the utilities are as in the perfect monitoring scenario even though their real utilities are affected by the noise.
Since the users select their actions based on their beliefs, once the pricing and the intervention rules are fixed, their interaction can still be modeled through the games (\ref{eq:gamma_P2}) and (\ref{eq:gamma_I2}), as in the perfect monitoring case. 
Analogously, the designer designs the pricing or the intervention rule based on its beliefs.
Hence, it has no reason to select rules different from the optimal (with respect to its beliefs) rules derived in Sections \ref{sec:per_pri} and \ref{sec:per_int}.
The only difference with respect to the perfect monitoring case is that the real performance of the system is different from the one expected by the users and the designer.

Notice that both users and designer might update their beliefs observing the real performance of the system. 
However, this might not be easy to do due to the lack of information.
On one hand, the designer designs an intervention rule and implements it in the intervention device, then it leaves the system. 
If the estimation errors are not correctly predicted in the design stage they affect the system, unless the designer implements a mechanism in the intervention device to reveal such errors. 
However, it might be difficult to discriminate between an estimation error and a real deviation of a user trying to increase its own utility.

On the other hand, the users might not be able to recognize the effect of an estimation error.
As an example, in the intervention scheme the estimation error triggers, occasionally, the intervention, with the consequent decrease of the throughput of a generic user $i$.
However, user $i$'s throughput decreases if another user increases its transmission probability as well.
Thus, $i$ is not able to understand if its utility has decreased due to the presence of the estimation errors or for some other reasons, and is not able to update its belief correctly.

This scenario has been considered in order to analyze how robust to an unknown noise the schemes derived in Sections \ref{sec:per_pri} and \ref{sec:per_int} are.

\subsection{The designer is aware of the estimation errors}\label{sec:imp_mon2}

In this scenario the users are not aware about the estimation noise, while the designer knows the distribution of the noise and knows that the users' beliefs are wrong.
Once the pricing and the intervention rules are given, the interaction between users can still be modeled through the games (\ref{eq:gamma_P2}) and (\ref{eq:gamma_I2}), in which the users act believing that their utilities are not affected by the estimation noise.
When designing the pricing or the intervention rule, the designer has to take into account both that the users act strategically, according to their mismatched perceptions,
and that the 
social welfare is affected by the noise.
It has to solve the \textbf{PD} and \textbf{ID} problems using the expectation of the noisy utilities given by Eq. (\ref{eq:Ui_P}) and (\ref{eq:Ui_I}) in the maximization and using the non-noisy utilities given by Eq. (\ref{eq:Ui_P2}) and (\ref{eq:Ui_I2}) in the constraints. 
In fact, the set of constraints represents the $NE$s of the game played by the users, in which the users select their action to maximize the utilities they believe to receive, i.e., the non-noisy utilities; while the maximization reflects the choice of the designer, that wants to maximize the real satisfaction of the users, represented by the expectation of the noisy utilities. 

Finally notice that, as described in Subsection \ref{sec:imp_mon1}, it might be difficult for the users to reveal the presence of the estimation errors by observing the real performance of the system.

\subsubsection{Pricing design}\label{sec:pri1}

Let $p_{k,1}$ denote the unique solution of equation 
\begin{equation}
\theta_k p_k^3 - \left( \theta_k + 4 \epsilon_k \sum_{i=1}^n \theta_i \right) p_k^2 + \left( 4 \epsilon_k \theta_k - \epsilon_k^2 \theta_k \right) p_k + \epsilon_k^2 \theta_k = 0
\label{eq:pk1}
\end{equation}
in $\left( 0 ,\, \epsilon_k \right)$, assuming $\dfrac{\theta_k}{\sum_{i=1}^n \theta_i} < \epsilon_k$.
Let $p_{k,2}$ denote the unique solution of equation 
\begin{equation}
-\theta_k p_k^3 + \left( \theta_k - 4 \epsilon_k \sum_{i=1}^n \theta_i \right) p_k^2 + \left( 4 \epsilon_k \theta_k + \left( 1 - \epsilon_k \right)^2 \theta_k \right) p_k - \left( 1 - \epsilon_k \right)^2 \theta_k = 0
\label{eq:pk1}
\end{equation}
in $\left( 1-\epsilon_k ,\, 1 \right)$, assuming $\dfrac{\theta_k}{\sum_{i=1}^n \theta_i} > 1 - \epsilon_k$.

In the proof of the following result, it is shown that $p_{k,1}$ and $p_{k,2}$ exist and are unique.

\begin{proposition}
The optimal unit price $c_k$ to adopt is, $\forall \, k \in \mathcal{N}$, 
 \begin{equation}
c_k =
\left\lbrace
\begin{array}{ll}
\dfrac{\theta_k}{p_{k,1}} & \mbox{if} \;\; \dfrac{\theta_k}{\sum_{i=1}^n \theta_i} < \epsilon_k \\
\sum_{i=1}^n \theta_i & \mbox{if} \;\; \epsilon_k \leq \dfrac{\theta_k}{\sum_{i=1}^n \theta_i} \leq 1 - \epsilon_k \\
\dfrac{\theta_k}{p_{k,2}} & \mbox{if} \;\; \dfrac{\theta_k}{\sum_{i=1}^n \theta_i} > 1-\epsilon_k \\
\end{array}
\right. 
\end{equation}
\label{prop:pric1}
\end{proposition}

\begin{IEEEproof}
See Appendix \ref{app:1}.
\end{IEEEproof}

\subsubsection{Intervention design}\label{sec:int1}

To design an intervention rule able to sustain the target action profile $\tilde{p}$, the designer has to satisfy the condition $r_k \geq \dfrac{1}{\tilde{p}_k}$, $\forall \, k$, provided by Lemma \ref{lem:2}.
The best option for the designer is to select $r_k = \dfrac{1}{\tilde{p}_k}$, $\forall \, k$, in order to sustain $\tilde{p}$ and, at the same time, to minimize the punishment adopted against $k$ when intervention is triggered by estimation errors.
Finally, the designer has to select the best $\tilde{p}_k$, for every user $k$.

Let $p_{k,3}$ denote the unique solution of equation 
\begin{equation}
\left( - \theta_k - \sum_{j=1}^n \theta_j \right) p_k^2 + \left( 2 \theta_k - 2 \epsilon_k \sum_{j=1}^n \theta_j \right) p_k + 2 \epsilon_k \theta_k = 0
\end{equation}
in $\left( 0 ,\, \epsilon_k \right)$.
In the proof of the following result, it is shown that $p_{k,3}$ exists and is unique.

\begin{proposition}
The optimal affine intervention rule to adopt is, for every user $k$, $r_k = \dfrac{1}{\tilde{p}_k}$ and 
\begin{equation}
\tilde{p}_k =
\left\lbrace
\begin{array}{ll}
p_{k,3} & \mbox{if} \;\; \epsilon_k > \dfrac{4 \theta_k}{4 \sum_{j=1}^n \theta_j - \sum_{j=1, j\neq k}^n \theta_j} \\
\dfrac{4 \theta_k + \epsilon_k \sum_{j=1, j\neq k}^n \theta_j}{4 \sum_{j=1}^n \theta_j} & \mbox{if} \;\; \epsilon_k \leq \dfrac{4 \theta_k}{4 \sum_{j=1}^n \theta_j - \sum_{j=1, j\neq k}^n \theta_j}
\end{array}
\right. 
\end{equation}
\label{prop:int}
\end{proposition}

\begin{IEEEproof}
See Appendix \ref{app:1_bis}.
\end{IEEEproof}

\subsection{Everybody is aware of the estimation errors}\label{sec:imp_mon3}

In this scenario both the designer and the users are aware of the estimation errors and they know their distribution.
The interaction between users must be modeled through the games (\ref{eq:gamma_P}) and (\ref{eq:gamma_I}) considering the real distribution of the noise in Eq. (\ref{eq:Ui_P}) and (\ref{eq:Ti_I}).
The designer has to solve the \textbf{PD} and \textbf{ID} problems using the utilities given by Eq. (\ref{eq:Ui_P}) and (\ref{eq:Ui_I}).

\subsubsection{Pricing design}\label{sec:pri2}

Once the pricing scheme is given, the interaction between users can be modeled with the game in Eq. (\ref{eq:gamma_P}),
where 
\begin{equation}
U_i^P(p) = \theta_i \ln \left[ p_i \prod_{j=1 , j \neq i}^n (1-p_j) \right] - \dfrac{c_i}{2 \epsilon_i} \int_{-\epsilon_i}^{\epsilon_i} \left[ p_i + x \right]_0^1  \partial x  
\label{eq:Ui_P3}
\end{equation}

Denote 
\begin{align}
&\mathcal{C} \left( \epsilon \right) = \left\lbrace x \; : \; \dfrac{1}{2} \leq x \leq 1-\epsilon \;\; \mbox{and} \;\; x \ln x - x \geq \dfrac{\epsilon}{4}-1 \right\rbrace \nonumber\\
&\overline{p}_k =
\left\lbrace
\begin{array}{ll}
-\dfrac{\epsilon_k}{2} + \dfrac{1}{2} \sqrt{\epsilon_k^2 + \dfrac{8 \epsilon_k \theta_k}{c_k}} & \mbox{if} \;\; \dfrac{\theta_k}{c_k} < \epsilon_k \\
\dfrac{\theta_k}{c_k} & \mbox{if} \;\; \epsilon_k \leq p_k \leq \dfrac{1}{2} \; \mbox{or} \; p_k \in \mathcal{C} \left( \epsilon_k \right) \\
1 & \mbox{otherwise}      
\end{array}
\right.
\label{eq:7}
\end{align}

\begin{lemma}
$\overline{p}_k$ is the unique $NE$ of the game $\Gamma^P$.
\label{lem:pric}
\end{lemma}

\begin{IEEEproof}
See Appendix \ref{app:2}.
\end{IEEEproof}

Consider the following notation:
\begin{align}
p_{k,4} &= \dfrac{\theta_k (2-\epsilon_k) }{4 \sum_i \theta_i} + \dfrac{1}{2} \sqrt{\left[ \dfrac{\theta_k (2-\epsilon_k)}{2 \sum_i \theta_i} \right]^2 + 4 \dfrac{\theta_k \epsilon_k}{2 \sum_i \theta_i}} \nonumber\\
p_{k,5} &= \max \mathcal{C} \left( \epsilon_k \right)
\end{align}

\begin{proposition}
The optimal unit price $c_k$ to adopt is, $\forall \, k \in \mathcal{N}$,
\begin{equation}
c_k =
\left\lbrace
\begin{array}{ll}
\dfrac{2 \epsilon_k \theta_k}{p_{k,4}  (p_{k,4} +\epsilon_k)} & \mbox{if} \;\; p_{k,4}  < \epsilon_k \\
\dfrac{\theta_k}{\epsilon_k} & \mbox{if} \;\; p_{k,4}  \geq \epsilon_k \;\; \mbox{and} \;\; \dfrac{\theta_k}{\sum_i \theta_i} \leq \epsilon_k \\
\sum_i \theta_i & \mbox{if} \;\; \epsilon_k \leq \dfrac{\theta_k}{\sum_i \theta_i} \leq \dfrac{1}{2} \;\; \mbox{or} \;\; \dfrac{\theta_k}{\sum_i \theta_i} \in \mathcal{C} \left( \epsilon_k \right) \\
\dfrac{\theta_k}{p_{k,5}} & \mbox{otherwise}      
\end{array}
\right.
\label{eq:8}
\end{equation}
\label{prop:pric2}
\end{proposition}

\begin{IEEEproof}
See Appendix \ref{app:3}.
\end{IEEEproof}

\subsubsection{Intervention design}\label{sec:int2}

Once the intervention scheme is given, the interaction between users can be modeled with the game in Eq. (\ref{eq:gamma_I}),
where 
\begin{align}
U_i^I(p) = \theta_i \ln \left[ p_i  \mathbb{E} \left[ \left[ r_i \left( p_i + n_i - \tilde{p}_i \right) \right]_0^1 \right] \prod_{j=1 , j \neq i}^n (1-p_j) \right] 
\label{eq:Ui_I3}
\end{align}

\begin{lemma}
Assume $2 \epsilon_k \leq \overline{p}_k \leq 1 - \epsilon_k $, $\overline{p}_k$ is the unique $NE$ of the game $\Gamma^I$ if $r_k \rightarrow + \infty $ and  $\tilde{p}_k = \overline{p}_k + \epsilon_k$.
\label{lem:int}
\end{lemma}

\begin{IEEEproof}
See Appendix \ref{app:4}.
\end{IEEEproof}

Lemma \ref{lem:int} states that, using an extreme rule, each user $k$ has the incentive to adopt a transmission probability $\overline{p}_k$ which is $\epsilon_k$ lower than $\tilde{p}_k$, to avoid the possibility of an intervention triggered by the estimation errors. 
This is true as long as $\tilde{p}_k$ is not too low, otherwise for user $k$ it is convenient to adopt a transmission probability closer to $\tilde{p}_k$, accepting the risk of an intervention triggered by the estimation errors.

\begin{proposition}
If $p_k^* = \dfrac{\theta_k}{\sum_{i=1}^n \theta_i} \geq 2 \epsilon_k $, for every user $k$, then the intervention rule $r_k \rightarrow + \infty $ and  $\tilde{p}_k = p_k^* + \epsilon_k$ is an optimal affine intervention.
\label{prop:2}
\end{proposition}

\begin{IEEEproof}
According to Lemma \ref{lem:int}, users have the incentive to adopt $p^* = \left(p_1^*, \ldots, p_n^* \right) $. 
In this case the intervention level is equal to $0$ because the estimation errors can not be higher than $\epsilon = \left(\epsilon_1^*, \ldots, \epsilon_n^* \right) $.
Thus, the outcome of the system is equal to the benchmark optimum.
Finally, this implies that $r_k \rightarrow + \infty $ and  $\tilde{p}_k = p_k^* + \epsilon_k$ define an optimal affine intervention rule, and, more specifically, also define an optimal intervention rule within the class of all intervention rules.
\end{IEEEproof}

\begin{corollary}
If $p_k^* = \dfrac{\theta_k}{\sum_{i=1}^n \theta_i} \geq 2 \epsilon_k $, the optimal affine intervention rule is optimal in the class of all intervention rules.
\label{cor:1}
\end{corollary}

We consider the following affine intervention rule, for every user $k$
\begin{align}
r_k &\rightarrow + \infty \nonumber\\
\tilde{p}_k &=
\left\lbrace
\begin{array}{ll}
p_k^*+\epsilon_k & \mbox{if} \;\; p_k^* \geq 2 \epsilon_k \\
3 \epsilon_k & \mbox{otherwise}      
\end{array}
\right.
\label{eq:con_int}
\end{align}

Eq. (\ref{eq:con_int}) defines an optimal intervention rule if $p_k^* \geq 2 \epsilon_k$, for every user $k$.
If $p_k^* < 2 \epsilon_k$, for some user $k$, the intervention rule might not be optimal. 
This rule is designed with the objective to minimize the intervention level.
In fact, each user $i$ has the incentive to adopt the action $\tilde{p}_i - \epsilon_i$, which results in an intervention level equal to $0$.

\subsection{Comparison between pricing and intervention and some results}\label{sec:res}

In the following we investigate how the social welfare and the total throughput vary increasing the number of users in the system, for the imperfect monitoring scenario, adopting both the pricing and the intervention schemes.
We consider the symmetric case, i.e., $\theta_i = \theta_j$ and $\epsilon_i = \epsilon_j$, $\forall \, i,j \in \mathcal{N}$.
Thus, the optimal transmission policy in the cooperative scenario, defined by Eq. (\ref{eq:p_opt}), is $p_k^* = \frac{1}{n}$, for every user $k$.

\begin{figure}
     \centering
          \includegraphics[width=\figw]{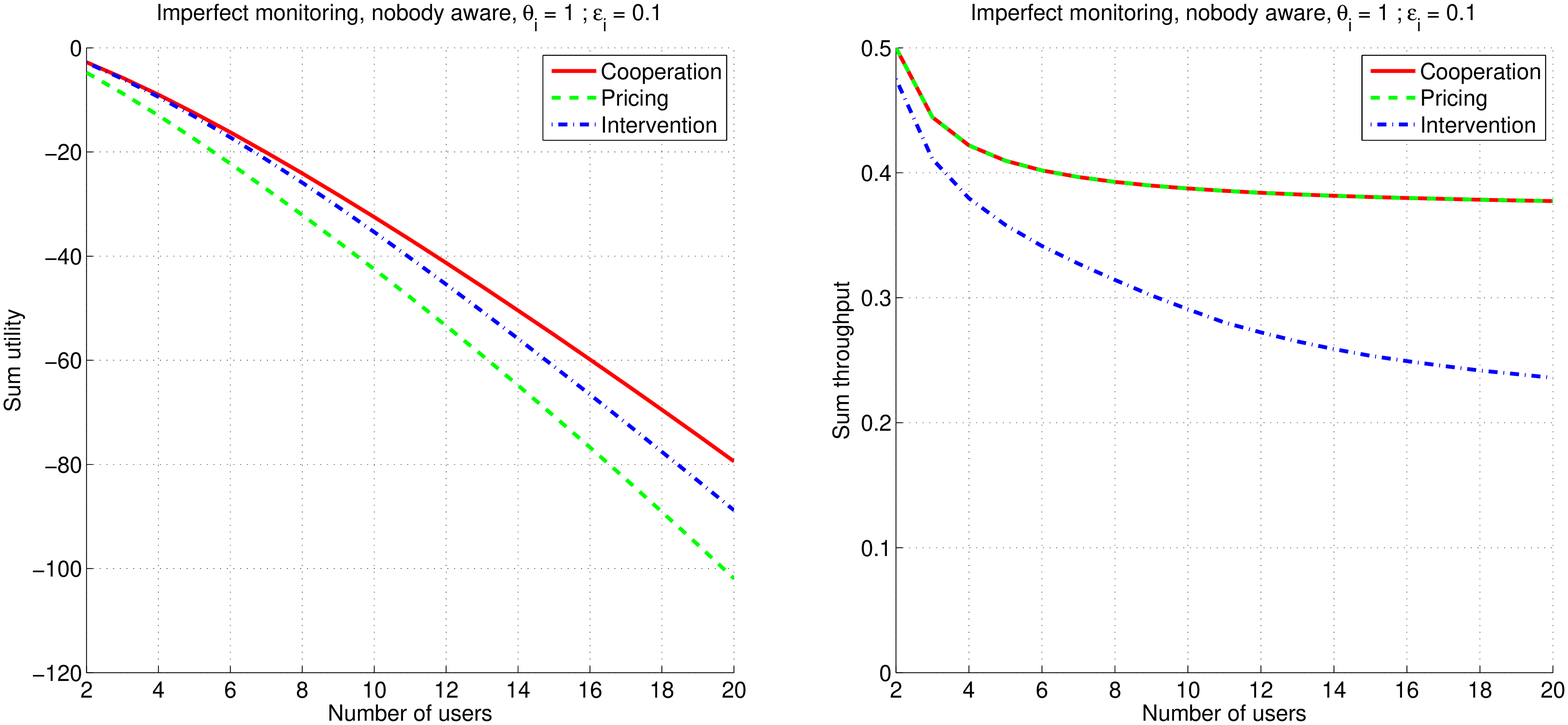}
\caption{Social welfare and total throughput vs. number of users, in the imperfect monitoring scenario, assuming nobody is aware of the estimation errors}
\label{fig:2}
\end{figure}

First we assume that nobody, neither the designer nor the users, is aware of the estimation errors.
As discussed in Subsection \ref{sec:imp_mon1}, the designer adopts the schemes derived in Section \ref{sec:per_mon} and the users, consequently, have the incentive to adopt the action $p_k^* = \frac{1}{n}$.
Fig \ref{fig:2} shows that the estimation errors have different effects in the two schemes. 
In the pricing scheme they do not affect the total throughput, and the social welfare is slightly affected only when the number of users exceeds $\frac{1}{\epsilon_i}=10$ (corresponding to the condition $p_k^* < \epsilon_i$).
In fact, if the number of users is less than or equal to $10$, each user is (on average) charged correctly.
Conversely, if the number of users exceeds $10$, the expectation of the estimated transmission probability $\hat{p}_k$ is higher than the real transmission probability $p_k^*$ and each user is (on average) slightly overcharged, resulting in a social welfare slightly lower than the one obtainable in the perfect monitoring scenario (see Fig. \ref{fig:1}).
In the intervention scheme the effect of the estimation errors is stronger. 
In fact, they occasionally trigger intervention, which decreases both the throughput and the utility experienced by each user.
Nevertheless, the social welfare adopting intervention is still higher than the social welfare adopting pricing.

\begin{figure}
     \centering
          \includegraphics[width=\figw]{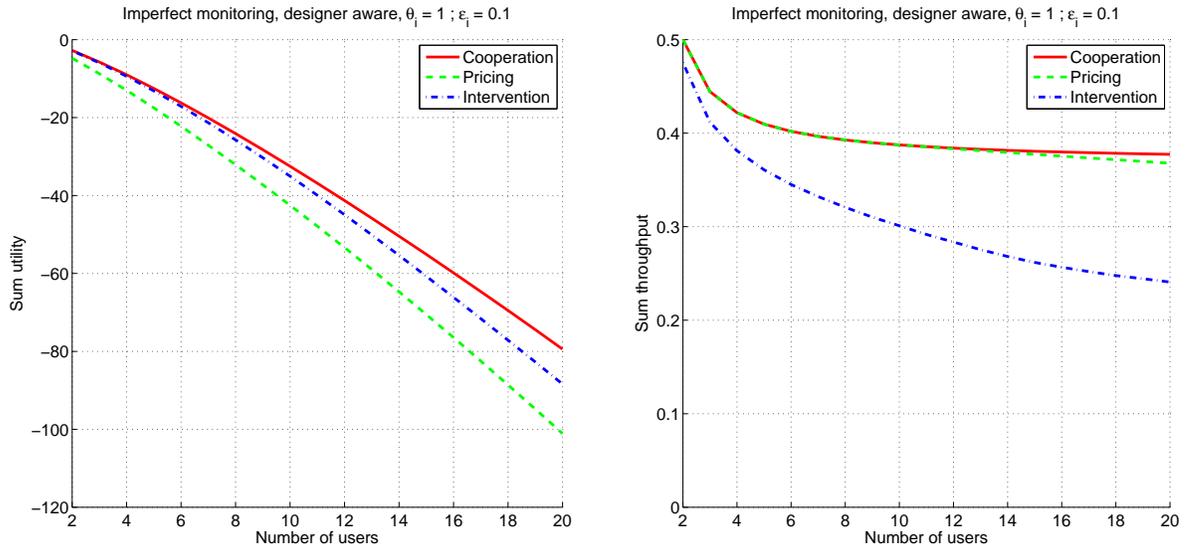}
\caption{Social welfare and total throughput vs. number of users, in the imperfect monitoring scenario, assuming that only the designer is aware of the estimation errors}
\label{fig:3}
\end{figure}

\begin{figure}
     \centering
          \includegraphics[width=\figw]{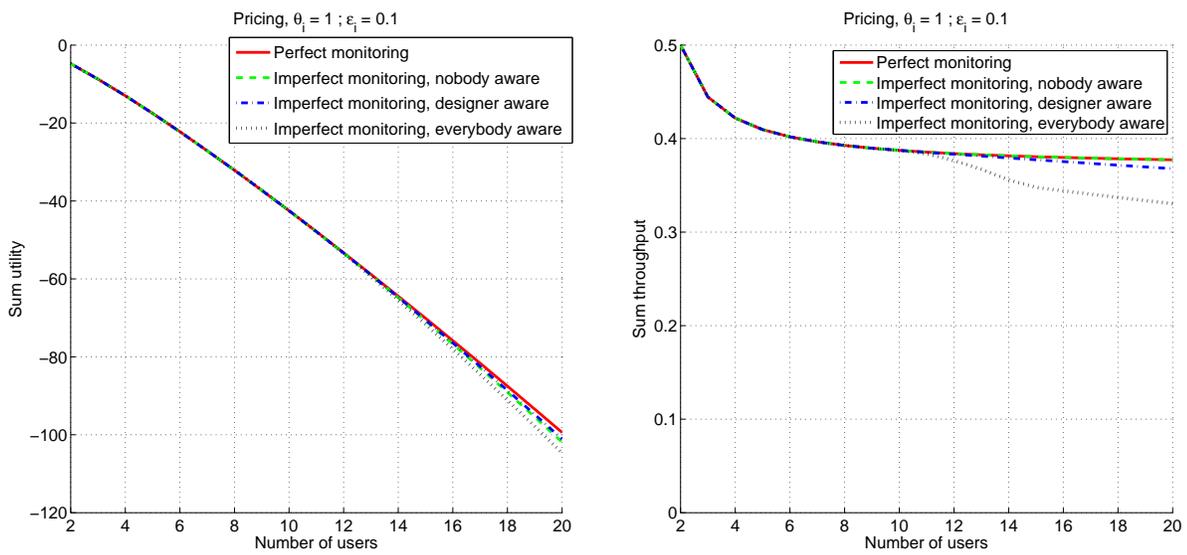}
\caption{Social welfare and total throughput vs. number of users adopting pricing, for different scenarios}
\label{fig:6}
\end{figure}

\begin{figure}
     \centering
          \includegraphics[width=\figw]{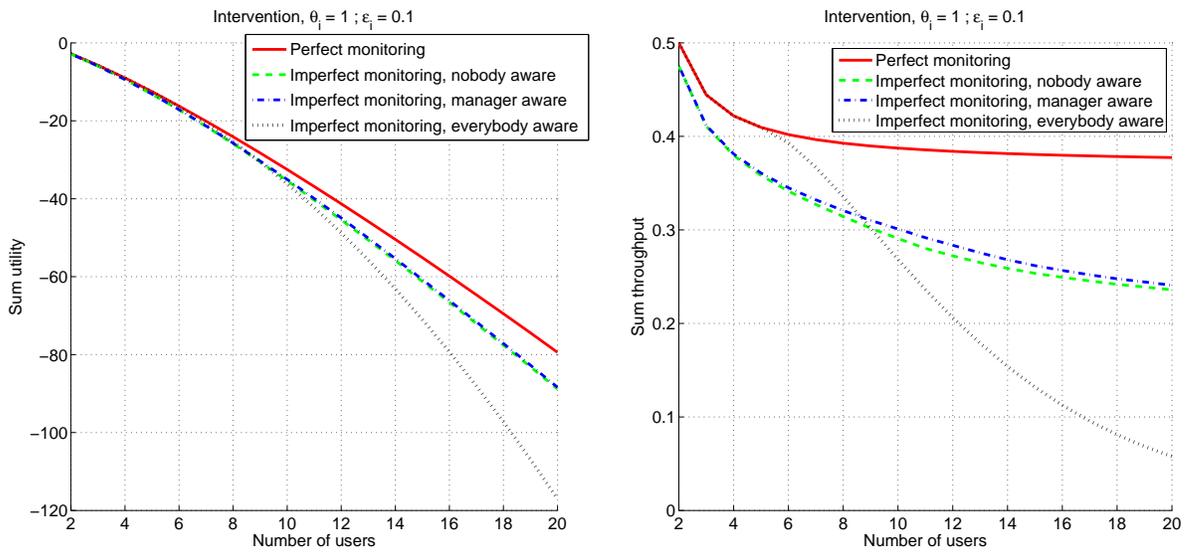}
\caption{Social welfare and total throughput vs. number of users adopting intervention, for different scenarios}
\label{fig:7}
\end{figure}

Now we consider the imperfect monitoring scenario assuming that only the designer is aware of the estimation errors.
In this case, the designer can adopt the optimal pricing and intervention schemes derived in Subsection \ref{sec:imp_mon2}.
The social welfares obtainable with both schemes, shown in Fig. \ref{fig:3}, are only slightly higher than the social welfares obtainable when nobody is aware of the estimation errors, shown in Fig. \ref{fig:2} (such differences will be clearer in Figs. \ref{fig:6} and \ref{fig:7}).
This means that the designer can not gain much with the additional information on the presence of estimation errors, and knowing their statistics.
In particular, for the pricing scheme such information is useless if the number of users is less than $10$, because the best pricing schemes derived in Subsections \ref{sec:imp_mon2} and \ref{sec:per_int} are identical in this situation.

\begin{figure}
     \centering
          \includegraphics[width=\figw]{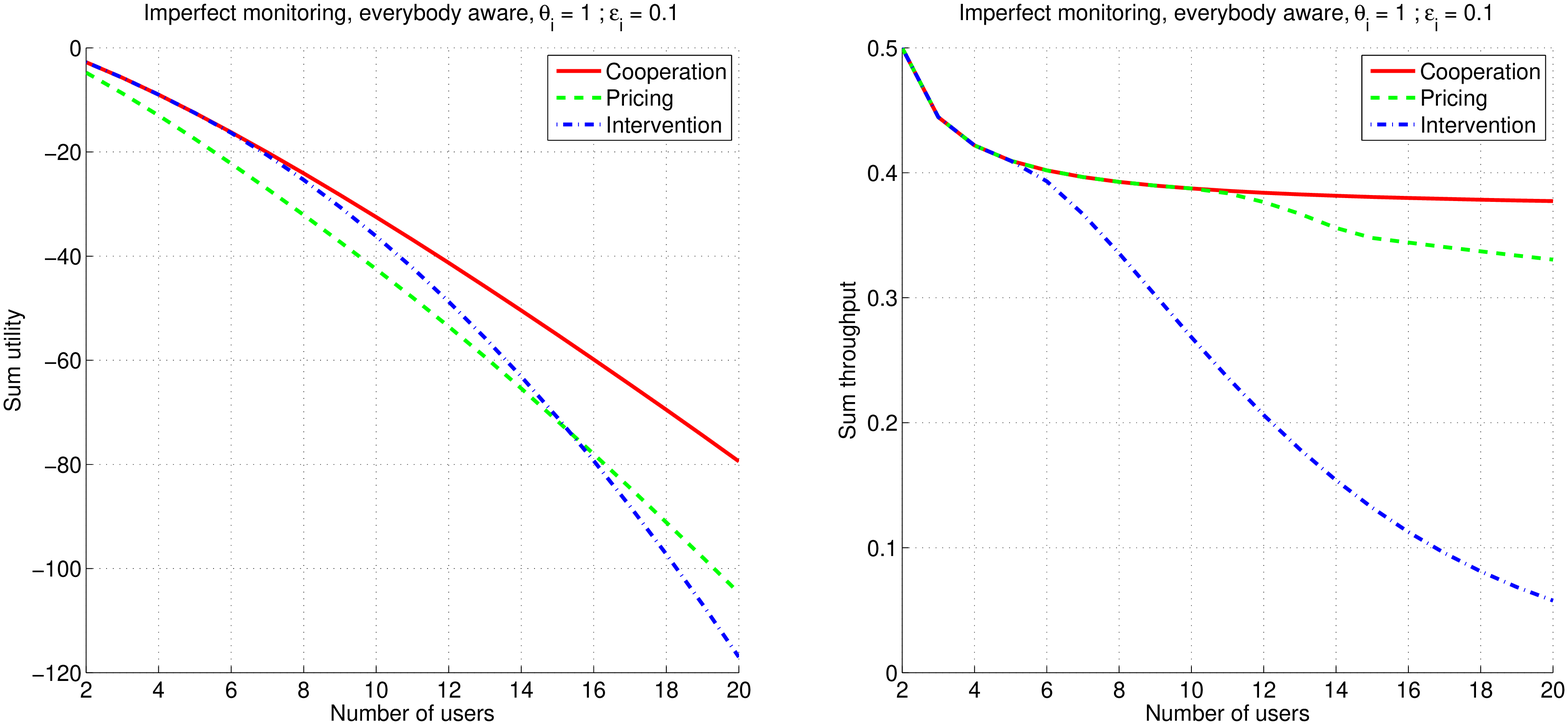}
\caption{Social welfare and total throughput vs. number of users, in the imperfect monitoring scenario, assuming everybody is aware of the estimation errors}
\label{fig:4}
\end{figure}

Now we investigate the performance achievable in the imperfect monitoring scenario assuming that everyone is aware of the estimation errors.
In this case the users, knowing that the noise might bias the payments (pricing) or the punishments (intervention), adopt a different $NE$ action profile.
Since the designer can foresee the users' behavior, it can adopt the pricing and intervention schemes derived in Subsection \ref{sec:imp_mon3}.
Fig. \ref{fig:4} shows that the performance attainable with the pricing scheme is very similar to the preceding cases, only slightly worse.
Conversely, the performance achievable with the intervention scheme is completely different from the preceding cases.
The intervention scheme is able to achieve the optimal social welfare as long as the number of users is less than or equal to $5$ (corresponding to the condition $p_k^* \geq 2 \epsilon_k$), as predicted by Proposition \ref{prop:2}). 
If the number of users is higher than $5$, both the total throughput and the social welfare decrease rapidly as the number of users increases. 
This trend is a consequence of the action adopted by the users in this situation, which is constant and equal to $2 \epsilon_k$ instead of scaling with the number of users.
This causes a rapid increase of the number of collisions.
Finally, this trend determines a threshold in the number of users such that, for a number of users lower than the threshold, intervention outperforms pricing, whereas, for a number of users higher than the threshold, pricing outperforms intervention.
The value of the threshold for the considered system parameters is equal to $15$.

\begin{figure}
     \centering
          \includegraphics[width=\figw]{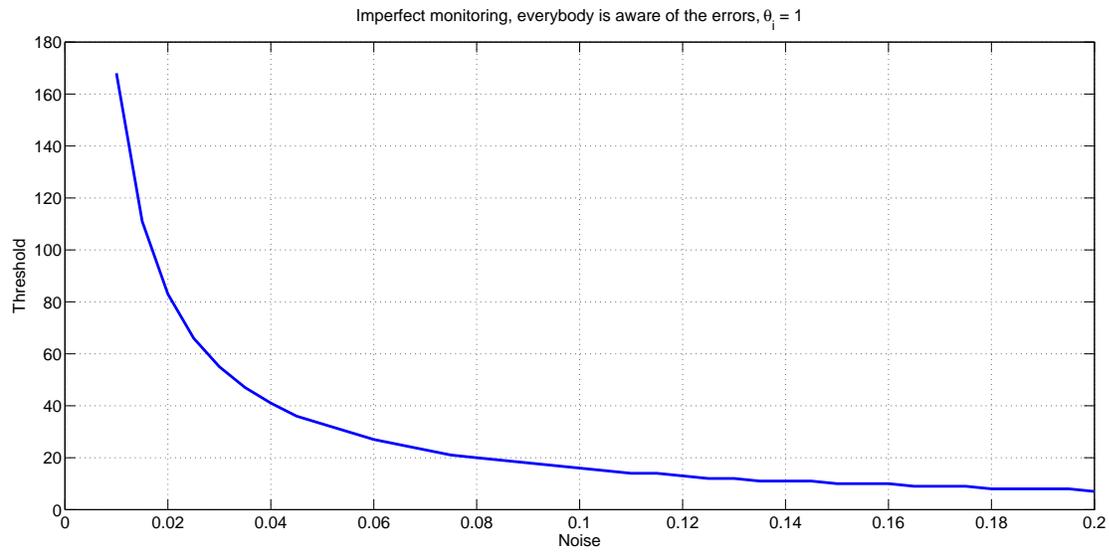}
\caption{Threshold vs. noise in the imperfect monitoring scenario, assuming everybody is aware of the estimation errors}
\label{fig:5}
\end{figure}

In Fig. \ref{fig:5} the value of the threshold is plotted varying $\epsilon_k$, the maximum intensity of the noise.
The threshold decreases as $\epsilon_k$ increases, because the intervention scheme is more sensitive to the estimation errors than the pricing scheme.
For the highest noise considered, i.e., $\epsilon_i = 0.2$, the intervention scheme outperforms the pricing scheme as long as the number of users is less than $9$.

In order to have a quantitative comparison between the different scenarios, in Figs. \ref{fig:6} and \ref{fig:7} we plot the social welfare and the total throughput  achievable for all considered cases, adopting pricing and intervention respectively.
In both Figures, we see that the system achieves the best performance if the monitoring is perfect. 
In case it is not, for the pricing scheme the best case is when only the designer is aware of the estimation errors, whereas the worst case is when also the users are aware of the estimation errors.
It is not surprising that, in a strategic setting, the more information the selfish users have the worse the efficiency of the equilibrium point. 
Conversely, for the intervention scheme we notice that when the users are aware of the estimation errors the social welfare might be higher than when they are not.
This result does not contradict the previous one, in fact it is caused by the additional information that the designer has as well: it knows that the users know that estimation errors exist, thus, it can design different intervention rules. 
In particular, it can adopt a more severe rule (e.g., the extreme rule, with $r_k \rightarrow + \infty$) that forces the users to keep their transmission probabilities low in order to avoid that the intervention is occasionally triggered by the estimation errors.
Fig. \ref{fig:7} shows that there is a threshold in the number of users such that, for a number of users lower than the threshold, it is socially convenient that the users are aware of the estimation errors, while for a number of users higher than the threshold it is not.

\begin{figure}
     \centering
          \includegraphics[width=\figw]{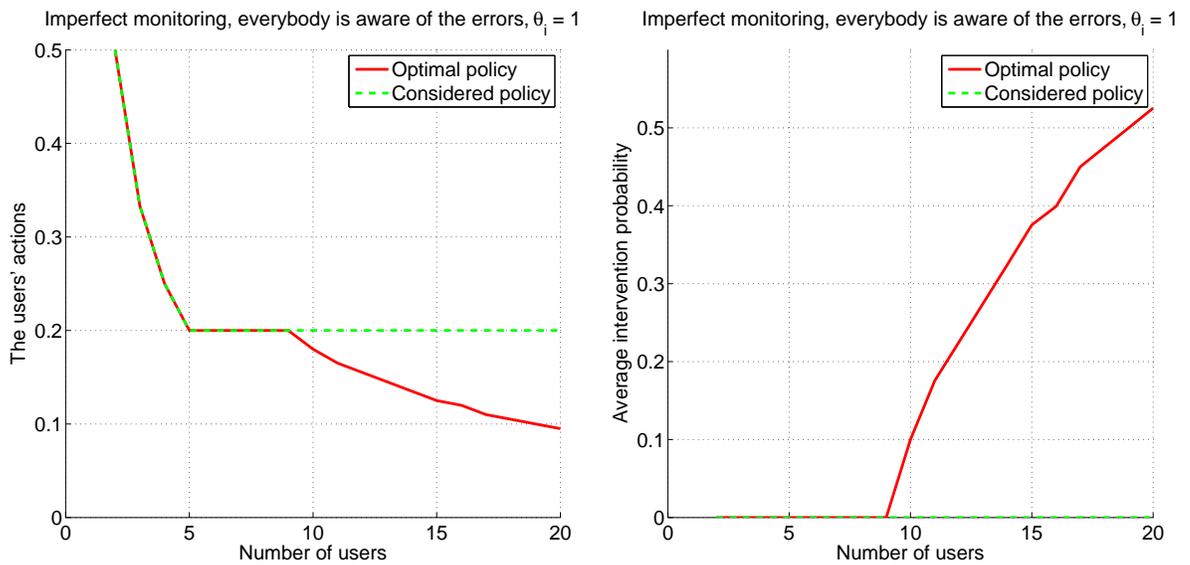}
\caption{The users' actions and the average level of intervention vs. number of users in the imperfect monitoring scenario, assuming everybody is aware of the estimation errors,  adopting the considered policy and the optimal one}
\label{fig:8}
\end{figure}

\begin{figure}
     \centering
          \includegraphics[width=\figw]{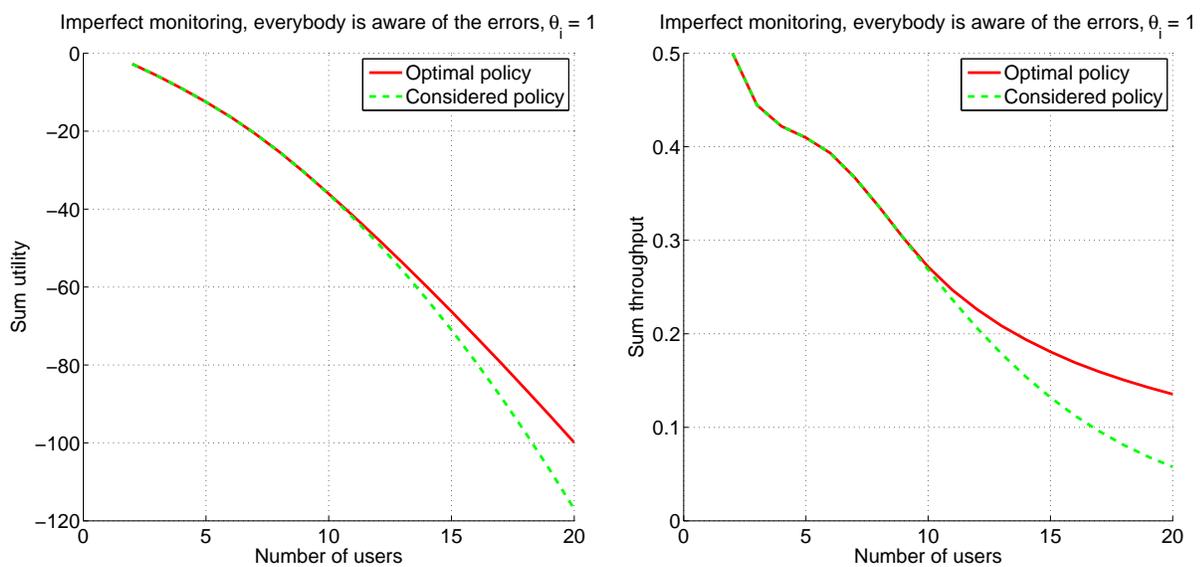}
\caption{Social welfare and total throughput vs. number of users in the imperfect monitoring scenario, assuming everybody is aware of the estimation errors,  adopting the considered policy and the optimal one}
\label{fig:9}
\end{figure}

Finally, in Figs. \ref{fig:8} and \ref{fig:9} we consider the imperfect monitoring scenario assuming that everyone is aware of the estimation errors, and we compare the considered intervention scheme of Eq. (\ref{eq:con_int}) with the optimal affine intervention rule.
The optimal affine intervention rule is computed adopting an exhaustive search algorithm.
Notice that this is possible because we consider a symmetric scenario. 
In asymmetric scenarios the calculation of the optimal rule through an exhaustive search algorithm would be computationally too expensive.
Fig. \ref{fig:8} shows the action selected by the users and the average intervention level varying the number of users, while Fig. \ref{fig:9} shows the social welfare and the total throughput varying the number of users.
Proposition \ref{prop:2} guarantees that the considered intervention rule is optimal for a number of users equal or lower than $5$ (corresponding to the condition $p_k^* \geq 2 \epsilon_k$).
However, as we can see, the considered intervention rule is optimal until $9$ users. 
If the number of users exceeds $9$, it is preferable to be more aggressive with the intervention rule, using a $\tilde{p}_k$ lower than $3 \epsilon_k$ and forcing the users to decrease their transmission probability as well, even though this means that the intervention is occasionally triggered.

\section{Conclusions}\label{sec:con}
In this paper we tackle the problem of designing pricing and intervention schemes to provide incentives for the users to exploit efficiently the channel resource in a contention game.
The design of the optimal schemes strongly depends on the parameters of the system, such as the statistics of the estimation errors, and on the information held by the designer and by the users.

In this work we have considered both the perfect monitoring and the imperfect monitoring scenarios, assuming, for the latter case, that (1) neither the designer nor the users are aware of the estimation errors, (2) only the designer is aware of the estimation errors, and (3) both the designer and the users are aware of the estimation errors.
The optimal linear pricing and affine intervention schemes have been analytically computed (for the case (3), the considered intervention scheme is optimal only in some conditions).

The analysis shows that the intervention scheme, differently from the pricing scheme, is able to achieve the optimal performance in the perfect monitoring scenario. 
On the other hand, in the imperfect monitoring scenario intervention might be triggered even when the users adopt the recommended actions, resulting in a degradation of the system performance.
Nevertheless, we noticed that intervention outperforms pricing in cases (1) and (2), while for case (3), as a rough general principle, intervention achieves greater efficiency than pricing when the number of users is small and the opposite is true when the number of users is large.

Another interesting result is related to the effect of the information held by the different entities.
While it is always desirable for the designer to have as much information as possible, the effect of the information held by the selfish users is not trivial.
In many cases it is preferable that the users are uninformed, but, sometimes, the information held by the users allows the designer to design better rules.
In our particular case, we have seen that the intervention can achieve the benchmark optimum if the users are aware of the estimation errors and the number of users is not too high. 
This suggests the idea, that might be true also in other settings, of hiding some system parameters from the users in determinate conditions.

Finally, the analysis in this paper can serve as a guideline for a designer to select between pricing and intervention and to design the best policy for the selected scheme, depending on some system parameters such as the number of users, the statistics of the monitoring noise and the information held by the designer and the users.

\appendices

\section{Proof of Proposition \ref{prop:pric1}}\label{app:1}

\begin{IEEEproof}

The unique $NE$ of the game (\ref{eq:gamma_P2}) is $p_i = \dfrac{\theta_i}{c_i}$, $i \in \mathcal{N}$.
Hence, the expected social welfare is
\begin{equation}
U(p) = \sum_{i=1}^n  \mathbb{E} \left[ \theta_i \ln \left[ p_i \prod_{j=1 , j \neq i}^n (1-p_j) \right] - c_i  \left[ p_i + \textit{n}_i \right]_0^1 \right] = \sum_{i=1}^n \theta_i \ln \left[ p_i \prod_{j=1 , j \neq i}^n (1-p_j) \right] - \dfrac{\theta_i}{p_i}  \mathbb{E} \left[ \left[ p_i + \textit{n}_i \right]_0^1 \right]
\label{eq:pk1}
\end{equation}
where, considering $\textit{n}_i \sim \mathcal{U} \left[ -\epsilon_i \, \epsilon_i \right] $,
\begin{equation}
\mathbb{E} \left[ \left[ p_i + \textit{n}_i \right]_0^1 \right] =
\left\lbrace
\begin{array}{ll}
\dfrac{\left( p_i + \epsilon_i \right)^2} {4 \epsilon_i} & \mbox{if} \;\; p_i < \epsilon_i \\
p_i & \mbox{if} \;\; \epsilon_i \leq p_i \leq 1 - \epsilon_i \\
\dfrac{-p_i^2 + 2(\epsilon_i+1)p_i + 2 \epsilon_i - \epsilon_i^2 - 1}{4 \epsilon_i} & \mbox{if} \;\; p_i > 1 - \epsilon_i
\end{array}
\right. 
\end{equation}

Therefore 
\begin{equation}
\dfrac{\partial U(p)}{\partial p_k} = 
\left\lbrace
\begin{array}{ll}
\dfrac{\theta_k}{p_k} - \dfrac{\sum_{i\neq k} \theta_i}{1-p_k} - \dfrac{\theta_k p_k^2 - \epsilon_k^2 \theta_k}{4 \epsilon_k p_k^2} & \mbox{if} \;\; p_k < \epsilon_k \\
\dfrac{\theta_k}{p_k} - \dfrac{\sum_{i\neq k} \theta_i}{1-p_k} & \mbox{if} \;\; \epsilon_k \leq p_k \leq 1 - \epsilon_k \\
\dfrac{\theta_k}{p_k} - \dfrac{\sum_{i\neq k} \theta_i}{1-p_k} + \theta_k \dfrac{p_k^2 - \left( 1 - \epsilon_k \right)^2 }{4 \epsilon_k p_k^2} & \mbox{if} \;\; p_k > 1 - \epsilon_k
\end{array}
\right. 
\end{equation}
and
\begin{equation}
\dfrac{\partial^2 U(p)}{\partial p_k^2} = 
\left\lbrace
\begin{array}{ll}
- \dfrac{\theta_k}{p_k^2} - \dfrac{\sum_{i\neq k} \theta_i}{\left( 1-p_k \right)^2} - \dfrac{\epsilon_k \theta_k}{2 p_k^3} & \mbox{if} \;\; p_k < \epsilon_k \\
- \dfrac{\theta_k}{p_k^2} - \dfrac{\sum_{i\neq k} \theta_i}{\left( 1-p_k \right)^2 } & \mbox{if} \;\; \epsilon_k \leq p_k \leq 1 - \epsilon_k \\
\dfrac{\theta_k}{p_k} - \dfrac{\sum_{i\neq k} \theta_i}{1-p_k} + \theta_k \dfrac{ \left( 1 - \epsilon_k \right)^2 }{2 \epsilon_k p_k^3} & \mbox{if} \;\; p_k > 1 - \epsilon_k
\end{array}
\right. 
\end{equation}
\begin{equation}
\dfrac{\partial^2 U(p)}{\partial p_k \partial p_j} = 0 \;\; , \;\; \forall \, j \neq k 
\end{equation}

$\dfrac{\partial^2 U(p)}{\partial p_k^2}$ is negative. 
For $p_k < \epsilon_k$ and $\epsilon_k \leq p_k \leq 1 - \epsilon_k$ this is trivial, for $p_k > 1 - \epsilon_k$ we have
\begin{align}
&\dfrac{\partial^2 U(p)}{\partial p_k^2} = \dfrac{- 2 \epsilon_k \theta_k p_k \left( 1 - p_k \right)^2 - 2 \epsilon_k \sum_{i\neq k} \theta_i p_k^3 + \left( 1 - \epsilon_k \right)^2 \theta_k \left( 1-p_k \right)}{2 \epsilon_k p_k^3 \left( 1 - p_k\right)^2 } < \nonumber\\
&< \dfrac{- 2 \epsilon_k \sum_{i\neq k} \theta_i p_k^3 + \left( 1 - \epsilon_k \right)^2 \theta_k \left( 1-p_k \right)}{2 \epsilon_k p_k^3 \left( 1 - p_k\right)^2 } <
\dfrac{- 2 \left( 1 - \epsilon_k \right)^3 \sum_{i\neq k} \theta_i + \epsilon_k \left( 1 - \epsilon_k \right)^2 \theta_k \left( 1-p_k \right)}{2 p_k^3 \left( 1 - p_k\right)^2 } < 0 \nonumber\\
\end{align}
where the second and third inequalities are valid because $\dfrac{\theta_k}{\sum_{i=1}^n \theta_i} > 1-\epsilon_k$ (as we will see, the optimal transmission probability $p_k$ is higher than $1-\epsilon_k$ if and only if $\dfrac{\theta_k}{\sum_{i=1}^n \theta_i}$ is higher than $1-\epsilon_k$) and $\epsilon_k \ll 1$ respectively.
Hence, the Hessian of $U(p)$ is negative definite (it is a diagonal matrix with strictly negative diagonal entries), so $U(p)$ is concave.
The global maximizer of $U(p)$ can be obtained with the first order condition.

For $p_k < \epsilon_k$ we obtain the condition
\begin{equation}
\theta_k p_k^3 - \left( \theta_k + 4 \epsilon_k \sum_{i=1}^n \theta_i \right) p_k^2 + \left( 4 \epsilon_k \theta_k - \epsilon_k^2 \theta_k \right) p_k + \epsilon_k^2 \theta_k = 0
\label{eq:eq}
\end{equation}
The solution of Eq. (\ref{eq:eq}) exists and is unique assuming $\dfrac{\theta_k}{\sum_{i=1}^n \theta_i} < \epsilon_k$. 
In fact the left hand side is a continuous function, decreasing in $p_k$ (its derivative with respect to $p_k$ corresponds to the second derivative of $U(p)$ with respect to $p_k$), equal to $\dfrac{\epsilon_k^2 \theta_k}{0^+} > 0$ for $p_k \rightarrow 0^+$ and to $\dfrac{\epsilon_k \sum_i \theta_i - \theta_k }{\epsilon_k \left( 1 - \epsilon_k \right) } < 0$ for $p_k \rightarrow \epsilon_k^-$.

For $\epsilon_k \leq p_k \leq 1 - \epsilon_k$ we obtain the condition
\begin{equation}
p_k = \dfrac{\theta_k}{\sum_{i=1}^n \theta_i} \rightarrow c_k = \sum_{i=1}^n \theta_i 
\end{equation}

For $p_k > 1-\epsilon_k$ we obtain the condition
\begin{equation}
-\theta_k p_k^3 + \left( \theta_k - 4 \epsilon_k \sum_{i=1}^n \theta_i \right) p_k^2 + \left( 4 \epsilon_k \theta_k + \left( 1 - \epsilon_k \right)^2 \theta_k \right) p_k - \left( 1 - \epsilon_k \right)^2 \theta_k = 0
\label{eq:eq2}
\end{equation}
The solution of Eq. (\ref{eq:eq2}) exists and is unique assuming $\dfrac{\theta_k}{\sum_{i=1}^n \theta_i} > 1-\epsilon_k$. 
In fact the left hand side is a continuous function, decreasing in $p_k$ (its derivative with respect to $p_k$ corresponds to the second derivative of $U(p)$ with respect to $p_k$), 
equal to $\dfrac{\theta_k - \left( 1-\epsilon_k \right) \sum_{i=1}^n \theta_i }{2 \epsilon_k + \left( 1-\epsilon_k \right)} > 0$ for $p_k \rightarrow \left( 1 - \epsilon_k \right)^+ $ and to $\dfrac{- \sum_{i\neq k} \theta_i }{0^+} < 0$ for $p_k \rightarrow 1^-$.

Finally, notice that the solutions found are consistent with the case considered and
\begin{align}
p_k < \epsilon_k \;\;\; &\Leftrightarrow \;\;\; \dfrac{\theta_k}{\sum_{i=1}^n \theta_i} < \epsilon_k \nonumber\\
\epsilon_k \leq p_k \leq 1 - \epsilon_k \;\;\; &\Leftrightarrow \;\;\; \epsilon_k \leq \dfrac{\theta_k}{\sum_{i=1}^n \theta_i} \leq 1 - \epsilon_k \nonumber\\
p_k > 1-\epsilon_k \;\;\; &\Leftrightarrow \;\;\; \dfrac{\theta_k}{\sum_{i=1}^n \theta_i} > 1-\epsilon_k 
\end{align}

\end{IEEEproof}

\section{Proof of Proposition \ref{prop:int}}\label{app:1_bis}

\begin{IEEEproof}

Given the intervention rule $\tilde{p}_k$ and $r_k = \dfrac{1}{\tilde{p}_k}$, $\forall \, k$, and the $NE$ action profile $p = \tilde{p}$, the intervention level for user $i$ is equal to
\begin{align}
f_i^I(\tilde{p}_i) = \left[ \dfrac{1}{\tilde{p}_i} ( \left[ \tilde{p}_i + n_i \right]_0^1 - \tilde{p}_i) \right]_0^1
\end{align}

Consequently, the expected throughput of a generic user $i$ and the social welfare are
\begin{align}
T_i^I(\tilde{p}) &= \mathbb{E} \left[ \tilde{p}_i \left( 1 - f_i^I(\tilde{p}_i) \right) \prod_{j=1 , j \neq i}^n (1-\tilde{p}_j) \right] = \tilde{p}_i \left( 1 - \mathbb{E} \left[ f_i^I(\tilde{p}_i) \right] \right) \prod_{j=1 , j \neq i}^n (1-\tilde{p}_j) \nonumber\\
U(\tilde{p}) &= \sum_{i=1}^n \theta_i \ln T_i(\tilde{p}) = 
\sum_{i=1}^n \theta_i \ln \left( \tilde{p}_i - \tilde{p}_i \mathbb{E} \left[ f_i^I(\tilde{p}_i) \right] \right) +  \sum_{i=1}^n \left( \sum_{j=1 , j \neq i}^n \theta_j \right) \ln \left( 1 - \tilde{p}_i \right) 
\end{align}

Now we want check if $U(\tilde{p})$ is concave analyzing its Hessian.
To do so, we first compute the average intervention level $\mathbb{E} \left[ f_i^I(\tilde{p}_i) \right]$, then we calculate $\dfrac{\partial U(\tilde{p})}{\partial \tilde{p}_i}$ and finally we compute $\dfrac{\partial^2 U(\tilde{p})}{\partial \tilde{p}_i^2}$ and $\dfrac{\partial^2 U(\tilde{p})}{\partial \tilde{p}_i \partial \tilde{p}_j}$, $i \neq j$.
Notice that, to do so, we should calculate each function in three different cases: for $\tilde{p}_i < \epsilon_i$, for $\epsilon_i \leq \tilde{p}_i \leq 1 - \epsilon_i$, and for $\tilde{p}_i > 1 - \epsilon_i$. 
However, to avoid a heavy notation, we do not take into consideration the case $\tilde{p}_i > 1 - \epsilon_i$. 
In fact this case is not interesting because, since $\epsilon_i \ll 1$, the best target action for user $i$ is close to $1$ if and only if there are few users in the network and the conditions are strongly asymmetric (i.e., $\theta_i \gg \theta_j$, $\forall \, j \neq i$).
On the contrary, we are interested in the case $\tilde{p}_i < \epsilon_i$ because the best $\tilde{p}_i$ scales with the number of users.
Thus, if the network is crowded, $\tilde{p}_i$ may become close to $0$. 

If $ \tilde{p}_i + n_i < 0 $ then $f_i^I(\tilde{p}_i) = 0$.
If $ \tilde{p}_i + n_i \geq 0 $ then $f_i^I(\tilde{p}_i) = \dfrac{1}{\tilde{p}_i} \left[ n_i \right]_0^{\tilde{p}_i} $.
Hence, we obtain
\begin{equation}
\mathbb{E} \left[ f_i^I(\tilde{p}_i) \right] =
\left\lbrace
\begin{array}{ll}
\dfrac{1}{2 \epsilon_i \tilde{p}_i} \int_0^{\tilde{p}_i} x \partial x + \dfrac{1}{2 \epsilon_i} \int_{\tilde{p}_i}^{\epsilon_i} \partial x = \dfrac{2 \epsilon_i - \tilde{p}_i}{4 \epsilon_i} & \mbox{if} \;\; \tilde{p}_i < \epsilon_i \\
\dfrac{1}{2 \epsilon_i \tilde{p}_i} \int_0^{\epsilon_i} x \partial x = \dfrac{\epsilon_i}{4 \tilde{p}_i} & \mbox{if} \;\; \tilde{p}_i \geq \epsilon_i 
\end{array}
\right. 
\end{equation}

Therefore 
\begin{equation}
\dfrac{\partial U(\tilde{p})}{\partial \tilde{p}_i} = 
\left\lbrace
\begin{array}{ll}
\dfrac{2 \theta_i \left( \epsilon_i + \tilde{p}_i \right)}{2 \epsilon_i \tilde{p}_i + \tilde{p}_i^2} - \dfrac{\sum_{j\neq i} \theta_j}{1-\tilde{p}_i} & \mbox{if} \;\; \tilde{p}_i < \epsilon_i \\
\dfrac{\theta_i}{\tilde{p}_i - \frac{\epsilon_i}{4} } - \dfrac{\sum_{j\neq i} \theta_j}{1-\tilde{p}_i} & \mbox{if} \;\; \tilde{p}_i \geq \epsilon_i 
\end{array}
\right. 
\end{equation}

\begin{equation}
\dfrac{\partial^2 U(\tilde{p})}{\partial \tilde{p}_i^2} = 
\left\lbrace
\begin{array}{ll}
\dfrac{2 \theta_i \left( - 2 \epsilon_i^2 - 2 \epsilon_i \tilde{p}_i - \tilde{p}_i^2 \right)}{\left( 2 \epsilon_i \tilde{p}_i + \tilde{p}_i^2 \right)^2 } - \dfrac{\sum_{j\neq i} \theta_j}{\left( 1-p_i \right)^2} & \mbox{if} \;\; \tilde{p}_i < \epsilon_i \\
\dfrac{- \theta_i}{\left( \tilde{p}_i - \frac{\epsilon_i}{4} \right)^2 } - \dfrac{\sum_{j\neq i} \theta_j}{\left( 1-\tilde{p}_i \right)^2 } & \mbox{if} \;\; \tilde{p}_i \geq \epsilon_i 
\end{array}
\right. 
\end{equation}

\begin{equation}
\dfrac{\partial^2 U}{\partial p_i \partial p_j} = 0 \;\; , \;\; \forall \, i \neq j 
\end{equation}

$\dfrac{\partial^2 U(\tilde{p})}{\partial \tilde{p}_i^2} < 0$. 
Hence, the Hessian of $U(\tilde{p})$ is negative definite (it is a diagonal matrix with strictly negative diagonal entries), so $U(\tilde{p})$ is concave.
The global maximizer of $U(\tilde{p})$ can be obtained with the first order condition, i.e., imposing $\frac{\partial U(\tilde{p})}{\partial \tilde{p}_i} = 0$.
Notice that $\frac{\partial U(\tilde{p})}{\partial \tilde{p}_i}$ is continuous, decreasing (because $\frac{\partial^2 U(\tilde{p})}{\partial \tilde{p}_i^2} < 0$), and tends to $+ \infty$ for $\tilde{p}_i \rightarrow 0^+$ and to $- \infty$ for $\tilde{p}_i \rightarrow 1^-$. 
Thus, there exists one and only one $\tilde{p}_i$ such that $\frac{\partial U(\tilde{p})}{\partial \tilde{p}_i} = 0$.

Imposing $\frac{\partial U(\tilde{p})}{\partial \tilde{p}_i} = 0$ for $\tilde{p}_i < \epsilon_i$, we obtain
\begin{equation}
\left( - \theta_i - \sum_{j=1}^n \theta_j \right) \tilde{p}_i^2 + \left( 2 \theta_i - 2 \epsilon_i \sum_{j=1}^n \theta_j \right) \tilde{p}_i + 2 \epsilon_i \theta_i = 0
\end{equation}

Imposing $\frac{\partial U(\tilde{p})}{\partial \tilde{p}_i} = 0$ for $\tilde{p}_i \geq \epsilon_i $, we obtain
\begin{equation}
\tilde{p}_i = \dfrac{4 \theta_i + \epsilon_i \sum_{j=1, j\neq i}^n \theta_j}{4 \sum_{j=1}^n \theta_j}
\end{equation}
This results is compatible with the condition $\tilde{p}_i \geq \epsilon_i $ if and only if
\begin{equation}
\epsilon_i \leq \dfrac{4 \theta_i}{4 \sum_{j=1}^n \theta_j - \sum_{j=1, j\neq i}^n \theta_j}
\end{equation}

\end{IEEEproof}

\section{Proof of Lemma \ref{lem:pric}}\label{app:2}

\begin{IEEEproof}

\begin{equation}
\mathbb{E} \left[ \left[ p_i + \textit{n}_i \right]_0^1 \right] =
\left\lbrace
\begin{array}{ll}
\dfrac{\left( p_i + \epsilon_i \right)^2} {4 \epsilon_i} & \mbox{if} \;\; p_i \leq \epsilon_i \\
p_i & \mbox{if} \;\; \epsilon_i < p_i \leq 1 - \epsilon_i \\
\dfrac{-p_i^2 + 2(\epsilon_i+1)p_i + 2 \epsilon_i - \epsilon_i^2 - 1}{4 \epsilon_i} & \mbox{if} \;\; p_i > 1 - \epsilon_i
\end{array}
\right. 
\end{equation}

\begin{equation}
U_i(p) = 
\left\lbrace
\begin{array}{ll}
\theta_i \ln \left[ p_i \prod_{j\neq i} (1-p_j) \right] - c_i \dfrac{(p_i + \epsilon_i)^2}{4 \epsilon_i} & \mbox{if} \;\; p_i < \epsilon_i \\
\theta_i \ln \left[ p_i \prod_{j\neq i} (1-p_j) \right] - c_i p_i & \mbox{if} \;\; \epsilon_i \leq p_i \leq 1 - \epsilon_i \\
\theta_i \ln \left[ p_i \prod_{j\neq i} (1-p_j) \right] - c_i \dfrac{-p_i^2 + 2(\epsilon_i+1)p_i + 2 \epsilon_i - \epsilon_i^2 - 1}{4 \epsilon_i} & \mbox{if} \;\; p_i > 1 - \epsilon_i \\
\end{array}
\right. 
\end{equation}

\begin{equation}
\dfrac{\partial U_i(p)}{\partial p_i} = 
\left\lbrace
\begin{array}{ll}
\dfrac{\theta_i}{p_i} - 2 c_i \dfrac{p_i + \epsilon_i}{4 \epsilon_i} & \mbox{if} \;\; p_i < \epsilon_i \\
\dfrac{\theta_i}{p_i} - c_i & \mbox{if} \;\; \epsilon_i \leq p_i \leq 1 - \epsilon_i \\
\dfrac{\theta_i}{p_i} - c_i \dfrac{-p_i + \epsilon_i + 1}{2 \epsilon_i} & \mbox{if} \;\; p_i > 1 - \epsilon_i \\
\end{array}
\right. 
\end{equation}

To compute the best response function of user $i$, we impose the first derivative of $U(p)$ equal to $0$ and we analyse the concavity of $U(p)$, with respect to $p_i$.
\begin{equation}
\dfrac{\partial U_i(p)}{\partial p_i} = 0 \longrightarrow p_i =
\left\lbrace
\begin{array}{ll}
\dfrac{-\epsilon_i}{2} + \dfrac{1}{2}\sqrt{\epsilon_i^2 + \dfrac{8 \epsilon_i \theta_i}{c_i}} & \mbox{if} \;\; \dfrac{\theta_i}{c_i} < \epsilon_i \\
\dfrac{\theta_i}{c_i} & \mbox{if} \;\; \epsilon_i \leq \dfrac{\theta_i}{c_i} \leq 1 - \epsilon_i \\
\dfrac{\epsilon_i + 1}{1} + \dfrac{1}{2}\sqrt{(\epsilon_i + 1)^2 - \dfrac{8 \epsilon_i \theta_i}{c_i}} & \mbox{if} \;\; \dfrac{1}{2} < \dfrac{\theta_i}{c_i} < 1 - \epsilon_i \\
\end{array}
\right. 
\end{equation}

\begin{equation}
\dfrac{\partial^2 U_i(p)}{\partial p_i^2} =
\left\lbrace
\begin{array}{ll}
- \dfrac{\theta_i}{p_i^2} - \dfrac{c_i}{2 \epsilon_i} & \mbox{if} \;\; p_i < \epsilon_i \\
\dfrac{\theta_i}{p_i^2} & \mbox{if} \;\; \epsilon_i \leq p_i \leq 1 - \epsilon_i \\
- \dfrac{\theta_i}{p_i^2} + \dfrac{c_i}{2 \epsilon_i} & \mbox{if} \;\; p_i > 1 - \epsilon_i \\
\end{array}
\right. 
\end{equation}

$\dfrac{\partial^2 U_i(p)}{\partial p_i^2} < 0$ for $p_i \in \left[ 0 ,\, \max \left( \sqrt{\dfrac{2 \epsilon_i \theta_i}{c_i}},\, 1-\epsilon_i \right) \right]$ and in $\max \left( \sqrt{\dfrac{2 \epsilon_i \theta_i}{c_i}}, \, 1-\epsilon_i \right)$ there is a change in the concavity. 
If $\dfrac{1}{2} < \dfrac{\theta_i}{c_i} \leq 1 - \epsilon_i$ then, after the change of concavity, the function reaches a local minimum in $p_i = \dfrac{\epsilon_i + 1}{2} + \dfrac{1}{2}\sqrt{(\epsilon_i + 1)^2 - \dfrac{8 \epsilon_i \theta_i}{c_i}}$ and then restarts to increase. 
Hence, in this case there are 2 local maxima: $p_i = \dfrac{\theta_i}{c_i}$ and $p_i = 1$. 
Comparing the $2$ maxima we obtain $U_i(\dfrac{\theta_i}{c_i}, p_{-i}) \geq U_i(1, p_{-1}) \Longleftrightarrow \dfrac{\theta_i}{c_i} \ln \dfrac{\theta_i}{c_i} - \dfrac{\theta_i}{c_i} \geq \dfrac{\epsilon_i}{4}-1$.

Summarizing:
\begin{enumerate}
\item[Case 1)] if $\dfrac{\theta_i}{c_i} < \epsilon_i$ then there is one local maximum which is the global maximum: $p_i = \dfrac{-\epsilon_i}{2} + \dfrac{1}{2} \sqrt{\epsilon_i^2 + \dfrac{8 \epsilon_i \theta_i}{c_i}}$
\item[Case 2)] If $\epsilon_i \leq \dfrac{\theta_i}{c_i} \leq \dfrac{1}{2}$ then there is one local maximum which is the global maximum: $p_i = \dfrac{\theta_i}{c_i}$
\item[Case 3)] If $\dfrac{1}{2} < \dfrac{\theta_i}{c_i} \leq 1 - \epsilon_i$ and $\dfrac{\theta_i}{c_i} \ln \dfrac{\theta_i}{c_i} - \dfrac{\theta_i}{c_i} \geq \dfrac{\epsilon_i}{4}-1$ then there are two local maxima and the global one is $p_i = \dfrac{\theta_i}{c_i}$
\item[Case 4)] If $\dfrac{1}{2} < \dfrac{\theta_i}{c_i} \leq 1 - \epsilon_i$ and $\dfrac{\theta_i}{c_i} \ln \dfrac{\theta_i}{c_i} - \dfrac{\theta_i}{c_i} < \dfrac{\epsilon_i}{4}-1$ then there are two local maxima and the global one is $p_i = 1$
\item[Case 5)] if $\dfrac{\theta_i}{c_i} > 1 - \epsilon_i$ then the function is increasing and the maximum is obtained for $p_i = 1$
\end{enumerate}

\end{IEEEproof}

\section{Proof of Proposition \ref{prop:pric2}}\label{app:3}

\begin{IEEEproof}

Considering that users adopt the $NE$ action profile (\ref{eq:7}), we want to maximize $U(p)$ with respect to $p_k$, $\forall k \, \in \mathcal{N}$.
The optimal $p_k$ must be lower than $1$, therefore we can consider only the first three cases listed at the end of Appendix \ref{app:2}.

We obtain:
\begin{equation}
\dfrac{\partial U(p)}{\partial p_k} = 
\left\lbrace
\begin{array}{ll}
\dfrac{\theta_k}{p_k} - \dfrac{\sum_{i\neq k} \theta_i}{1-p_k} + \dfrac{\theta_k \epsilon_k}{2 p_k^2} & \mbox{if} \;\; p_k < \epsilon_k \\
\dfrac{\theta_k}{p_k} - \dfrac{\sum_{i\neq k} \theta_i}{1-p_k} & \mbox{if} \;\; \epsilon_k \leq p_k \leq p_{k,5} \\
\end{array}
\right. 
\end{equation}

\begin{equation}
\dfrac{\partial U(p)}{\partial p_i} = 0 \longrightarrow p_i =
\left\lbrace
\begin{array}{ll}
p_{k,4}  & \mbox{if} \;\; p_k < \epsilon_k \\
\dfrac{\theta_k}{\sum_{i\neq k} \theta_i} & \mbox{if} \;\; \epsilon_k \leq p_k \leq p_{k,5} \\
\end{array}
\right. 
\end{equation}

\begin{equation}
\dfrac{\partial^2 U(p)}{\partial p_k^2} =
\left\lbrace
\begin{array}{ll}
\dfrac{- \theta_k}{p_k^2} - \dfrac{\sum_{i\neq k} \theta_i}{(1-p_k)^2} - \dfrac{\theta_k \epsilon_k}{p_k^3} & \mbox{if} \;\; p_k < \epsilon_k \\
\dfrac{- \theta_k}{p_k^2} - \dfrac{\sum_{i\neq k} \theta_i}{(1-p_k)^2} & \mbox{if} \;\; \epsilon_k \leq p_k \leq p_{k,5} \\
\end{array}
\right. 
\end{equation}

\begin{equation}
\dfrac{d^2 G(p)}{d p_k dp_i} = 0 \; , \; i \neq k 
\end{equation}

The Hessian of $U(p)$ is negative definite in $[0, \, p_{k,5} ]$. $U(p)$ is a continuous and concave function in $[0, \, p_{k,5} ]$, increasing in $p_k = 0$. 
However, its first partial derivative is not continuous in $p_k = \epsilon_k$. 
In particular, if there exists a user $k$ such that $\dfrac{\partial G(p)}{\partial p_k} \neq 0$ in $[0, \, p_{k,5} ]$, then either (1) $\dfrac{\partial U(p)}{\partial p_k} > 0$ for $p_k < \epsilon_k$ and $\dfrac{\partial U(p)}{\partial p_k} < 0$ for $p_k > \epsilon_k$, or (2) $U(p)$ increases in $p_k$ until reaching a maximum in $p_k = p_{k,5} $. 
Finally, the global maximum is located where partial derivatives are equal to $0$ or, in case this condition is not satisfied for some users $k$, in $p_k = \epsilon_k$ if $\dfrac{\partial G(p)}{\partial p_k} > 0$ for $p_k < \epsilon_k$ and $\dfrac{\partial G(p)}{\partial p_k} < 0$ for $p_k > \epsilon_k$, or in $p_{k,5} $ otherwise; i.e.,

\begin{equation}
p_k =
\left\lbrace
\begin{array}{ll}
p_{k,4} & \mbox{if} \;\; p_{k,4} < \epsilon_k \\
\epsilon_k & \mbox{if} \;\; p_{k,4} \geq \epsilon_k \;\; \mbox{and} \;\; \dfrac{\theta_k}{\sum_i \theta_i} \leq \epsilon_k \\
\dfrac{\theta_k}{\sum_i \theta_i} & \mbox{if} \;\; \epsilon_k \leq \dfrac{\theta_k}{\sum_i \theta_i} \leq \dfrac{1}{2}  \;\; \mbox{or} \;\; \dfrac{\theta_k}{\sum_i \theta_i} \in \mathcal{C} \left( \epsilon_k \right) \\
p_{k,5} & \mbox{otherwise}      
\end{array}
\right.
\label{eq:9}
\end{equation}
which is equivalent to Eq. (\ref{eq:8}).

\end{IEEEproof}

\section{Proof of Lemma \ref{lem:int}}\label{app:4}

\begin{IEEEproof}

We study $i$'s utility, $U_i^I(p)$, varying $i$'s action, $p_i$.
To do so, we first analyze the average intervention level $ E := \mathbb{E} \left[ \left[ r_i \left( \left[ p_i + n_i \right]_0^1 - \overline{p}_i - \epsilon_i \right) \right]_0^1 \right]$, for $r_i \rightarrow +\infty$.

If $p_i < \overline{p}_i$, the term that multiplies $r_i$ is always negative (notice that $\left[ p_i + n_i \right]_0^1 \leq p_i + \epsilon_i $) and, consequently, the intervention level is always equal to $0$ and $E = 0$.

If $p_i > \overline{p}_i + 2 \epsilon_i$, the term that multiplies $r_i$ is always positive (notice that $\left[ p_i + n_i \right]_0^1 \geq p_i - \epsilon_i$) and, consequently, the intervention level is always equal to $1$ and $E = 1$.

If $ \overline{p}_i \leq p_i \leq \overline{p}_i + 2 \epsilon_i$, the intervention might be $0$ or $1$, depending on the value of the estimation error $n_i$. 
Notice that, in this case, $p_i + n_i \geq 0$.
Thus, whenever $n_i$ is higher than $\overline{p}_i + \epsilon_i - p_i$, the intervention is $1$, and the average intervention level is equal to
\begin{equation}
E =  \dfrac{1}{2 \epsilon_i} \int_{\overline{p}_i + \epsilon_i - p_i }^{\epsilon_i } \partial x = \dfrac{1}{2 \epsilon_i} \left( p_i - \overline{p}_i \right) 
\end{equation}

Hence, we obtain
\begin{equation}
U_i(p) = 
\left\lbrace
\begin{array}{ll}
\theta_i \ln p_i + \theta_i \ln \left[ \prod_{j \neq i} \left( 1 - p_j \right) \right] & \mbox{if} \;\; p_i < \overline{p}_i \\
\theta_i \ln \left[ p_i \left( 1 - \dfrac{1}{2 \epsilon_i} \left( p_i - \overline{p}_i \right) \right) \right] + \theta_i \ln \left[ \prod_{j \neq i} \left( 1 - p_j \right) \right] & \mbox{if} \;\; \overline{p}_i \leq p_i \leq \overline{p}_i + 2 \epsilon_i \\
- \infty & \mbox{if} \;\; p_i > \overline{p}_i + 2 \epsilon_i 
\end{array}
\right. 
\end{equation}

To predict the best action for user $i$, we study the trend of $U_i(p)$ varying $p_i$ in the interval $\left[ 0, \overline{p}_i + 2 \epsilon_i \right) $.
To do so, we calculate $\dfrac{\partial U_i(p)}{\partial p_i}$ and $\dfrac{\partial^2 U_i(p)}{\partial p_i^2}$. and we study their sign.

\begin{equation}
\dfrac{\partial U_i(p)}{\partial p_i} = 
\left\lbrace
\begin{array}{ll}
\dfrac{\theta_i}{p_i} & \mbox{if} \;\; p_i < \overline{p}_i \\
\theta_i \dfrac{1 + \frac{1}{2 \epsilon_i} \overline{p}_i - \frac{1}{\epsilon_i} p_i}{\left( 1 + \frac{1}{2 \epsilon_i} \overline{p}_i \right) p_i - \frac{1}{2 \epsilon_i} p_i^2 } & \mbox{if} \;\; \overline{p}_i \leq p_i \leq \overline{p}_i + 2 \epsilon_i
\end{array}
\right. 
\end{equation}

\begin{equation}
\dfrac{\partial^2 U_i(p)}{\partial p_i^2} = 
\left\lbrace
\begin{array}{ll}
- \dfrac{\theta_i}{p_i^2} & \mbox{if} \;\; p_i < \overline{p}_i \\
\theta_i \dfrac{ \frac{1}{\epsilon_i} \left( p_i - \overline{p}_i \right) - 1 + \frac{p_i}{2 \epsilon_i^2} \left( \overline{p}_i - p_i \right) - \frac{p_i^2}{2 \epsilon_i^2} - \frac{\overline{p}_i^2}{4 \epsilon_i^2} } {\left[ \left( 1 + \frac{1}{2 \epsilon_i} \overline{p}_i \right) p_i - \frac{1}{2 \epsilon_i} p_i^2 \right]^2 } & \mbox{if} \;\; \overline{p}_i \leq p_i \leq \overline{p}_i + 2 \epsilon_i
\end{array}
\right. 
\end{equation}

$\dfrac{\partial^2 U_i(p)}{\partial p_i^2} < 0$ for $\overline{p}_i \leq p_i \leq \overline{p}_i + 2 \epsilon_i$.
In fact, for $\overline{p}_i \leq p_i \leq \overline{p}_i + \epsilon_i$
\begin{equation}
\dfrac{1}{\epsilon_i} \left( p_i - \overline{p}_i \right) - 1 + \dfrac{p_i}{2 \epsilon_i^2} \left( \overline{p}_i - p_i \right) - \dfrac{p_i^2}{2 \epsilon_i^2} - \dfrac{\overline{p}_i^2}{4 \epsilon_i^2} \leq 1 - 1 + 0 \dfrac{p_i^2}{2 \epsilon_i^2} - \dfrac{\overline{p}_i^2}{4 \epsilon_i^2} \leq 0 
\end{equation}
For $\overline{p}_i + \epsilon_i \leq p_i \leq \overline{p}_i + 2 \epsilon_i$
\begin{align}
&\dfrac{1}{\epsilon_i} \left( p_i - \overline{p}_i \right) - 1 + \dfrac{p_i}{2 \epsilon_i^2} \left( \overline{p}_i - p_i \right) - \frac{p_i^2}{2 \epsilon_i^2} - \dfrac{\overline{p}_i^2}{4 \epsilon_i^2} \leq 2 - 1 -\dfrac{p_i}{2 \epsilon_i} - \frac{p_i^2}{2 \epsilon_i^2} \leq 1 - \dfrac{\overline{p}_i + \epsilon_i }{2 \epsilon_i} - \dfrac{\left( \overline{p}_i + \epsilon_i \right)^2} {2 \epsilon_i^2} = \nonumber\\
&\quad \quad \quad = 1 - \dfrac{1}{2} - \dfrac{\overline{p}_i }{2 \epsilon_i} - \dfrac{\overline{p}_i^2 + 2 \overline{p}_i \epsilon_i + \epsilon_i^2 }{2 \epsilon_i^2} = - \dfrac{\overline{p}_i }{2 \epsilon_i} - \dfrac{\overline{p}_i^2 }{2 \epsilon_i^2} - \dfrac{\overline{p}_i }{\epsilon_i} \leq 0
\end{align}

Thus, $\dfrac{\partial U_i(p)}{\partial p_i}$ is decreasing in $\left[ \overline{p}_i, \overline{p}_i + 2 \epsilon_i\right] $.
Since $\dfrac{\partial U_i(p)}{\partial p_i} > 0$ in $\left[ 0, \overline{p}_i, \right) $, a necessary and sufficient condition such that $\overline{p}_i$ is a global maximum is that $\dfrac{\partial U_i(p)}{\partial p_i} \leq 0$ for $p_i \rightarrow \overline{p}_i^+$.
Imposing such a condition we obtain $\overline{p}_i \geq 2 \epsilon_i$, which concludes the proof.

\end{IEEEproof}

\bibliographystyle{IEEEtran}
\bibliography{IEEEabrv,bibcok}

\end{document}